# Enhancement of the Solar Water Splitting Efficiency Mediated by Surface Segregation in Ti-doped Hematite Nanorods

*Stefan Stanescu[a]\*, Théo Alun[b], Yannick J. Dappe[b], Dris Ihiawakrim[c], Ovidiu Ersen[c], and Dana Stanescu[b]\**

[a] Synchrotron SOLEIL, L'Orme des Merisiers, Départementale 128, 91190 Saint-Aubin, France

E-mail: stefan.stanescu@synchrotron-soleil.fr

[b] SPEC, CEA, CNRS, Université Paris-Saclay, CEA Saclay 91191 Gif-sur-Yvette Cedex, France

E-mail: dana.stanescu@cea.fr

[c] Institut de Physique et Chimie des Matériaux de Strasbourg (IPCMS), CNRS UMR 7504, 23 rue du Loess, BP43, 67034 Strasbourg, France

ABSTRACT

Band engineering is employed thoroughly and targets technologically scalable photoanodes for solar water splitting applications. Complex and costly recipes are necessary, often for average performances. Here we report simple photoanode growth and thermal annealing, with effective band engineering results. By comparing Ti-doped hematite photoanodes annealed under Nitrogen to photoanodes annealed in air, we found strongly enhanced photocurrent, of more than 200 % in the first case. Using electrochemical impedance spectroscopy and synchrotron X-rays spectromicroscopies we demonstrate that oxidized surface states and increased density of charge carriers are responsible for the enhanced photoelectrochemical activity. Surface states are found to be related to the formation of pseudo-brookite clusters by surface Ti segregation. Spectro-ptychography is used for the first time at Ti $L_3$ absorption edge to isolate Ti chemical coordination arising from pseudo-brookite clusters contribution. Correlated with electron microscopy investigation and Density Functional Theory (DFT) calculations, the synchrotron spectromicroscopy data prove unambiguously the origin of the better photoelectrochemical activity of $N_2$-annealed Ti-doped hematite nanorods. Finally, we present here a handy and cheap surface engineering method, beyond the known oxygen vacancy doping, allowing a net gain in the photoelectrochemical activity for the hematite-based photoanodes.

## 1. INTRODUCTION

Hydrogen production by solar water splitting (SWS) using abundant and eco-friendly photoelectrode materials is very appealing nowadays. The overall efficiency of SWS reaction is directly related to the photoelectrochemical (PEC) activity including both oxidation and reduction reactions that occur at the interface between photoelectrodes and the aqueous electrolyte. Hematite is the archetype semiconducting material used as photoanode, presenting a band gap of 2.15 eV, perfectly matching the solar spectrum for optimized absorption and thus for direct SWS applications. Compared to state-of-the-art oxide semiconductors[1] used for solar water splitting, pristine hematite presents low efficiency because of the reduced hole mean free path (~ 2 – 4 nm) and poor surface kinetics related to a complex oxygen evolution reaction (OER).[2-6] It was demonstrated that surface kinetics during SWS can be tuned by controlling the surface states nature using various approaches: doping,[7-9] hetero- and nano-structuring,[10-14] annealing,[15,16] catalyst film coating,[17-22] etc. In particular the synergic effect between Ti-doping and induced oxygen vacancies through thermal treatments during or post growth under oxygen depleted atmosphere has shown its efficiency.[23-27] Various mechanisms were suggested to contribute to the enhancement of PEC activity. Zhao et al.[24] have shown that by annealing pristine and Ti-doped hematite photoanodes in Nitrogen gas at 600°C the photocurrent is increased by 200 % and 67 % respectively. They assigned this enhancement to oxygen vacancies generated by nitrogen treatment, leading to carrier density increase. Opposite behaviors were found for the charge transfer processes, comparing pristine hematite and Ti-doped hematite, suggesting increased transfer resistance for the later one. To the contrary, Wang et al.[26] demonstrated strong electrocatalytic surface contribution of the induced oxygen vacancies along to the expected improvement of the bulk conductivity. Such divergences arise from the variety of the employed sample preparation methods, leading to different sensitivities to subtle surface effects. These conclusions are obtained from PEC macroscopic measurements only, nanoscale information being absent. A perfect demonstration is given by Zhang et al.,[23] in the case of highly ordered attached Ti-modified hematite mesocrystals. Their study presents strongly enhanced PEC properties promoted by a double contribution: first, the formation of interfacial oxygen vacancies yielding high carrier density, and second, shorter depletion width (< 10 nm) over large regions through formation of rutile $TiO_2$ at the mesocrystals surface, as determined by high resolution electron microscopy. Indeed, complex hematite-Ti based heterostructures ($Fe_2O_3/Fe_2TiO_5$ or $Fe_2O_3/FeTiO_3$) are known to be responsible for PEC activity enhancement[19,28-31] owing to strongly increased surface charge separation. This leads to increased number of holes injection at the interface with the electrolyte.

Cumbersome experimental strategies are needed to obtain complex heterostructures, necessitating several growth or treatment cycles and involving various chemical processes. Here, we report on simple heterostructuring owing to surface segregation of a Ti-rich phase during the annealing under Nitrogen, used to transform the initial Ti-doped akaganeite phase obtained by aqueous growth[15] into Ti-doped hematite. We recorded strong enhancement of the photocurrent and lowering of the flat-band values for the Nitrogen-annealed photoanodes compared to air-annealed ones. It is important to mention that the absolute photocurrent values we determined are lower compared to cutting-edge ones reported for hematite-based photoanodes,[32] even though not strictly in the same working conditions. Our work aims to compare strictly the PEC activity between two kinds of hematite-based photoanodes, air- and Nitrogen-annealed, and to clarify the origin of the enhancement observed for the latter. For this purpose, we correlate macroscopic PEC (photo-voltammetry, EIS – Electrochemical Impedance Spectroscopy) and state-of-the-art synchrotron X-rays spectromicroscopy measurements (S-XPEEM – Shadow X-rays Photoemission Electron Microscopy, STXM – Scanning Transmission X-rays Microscopy, spectro-ptychography), completed by electron microscopy (STEM – Scanning Transmission Electron Microscopy). The results demonstrate the formation of surface pseudo-brookite clusters yielding partial Ti:$Fe_2O_3/Fe_2TiO_5$ interface. Our data suggest that this heterostructuring is at the origin of the better charge separation through the formation of corresponding surface states. Interestingly, we observe that a very low fraction (< 5 %) of the Ti-doped hematite is covered by pseudo-brookite clusters and thus responsible for the enhanced PEC activity. Efficient charge transport is ensured by enhanced bulk conduction promoted by *n*-type conduction through Ti and Oxygen vacancy doping. Finally, we show subtle differences of chemical and electronic structures of hematite nanorods, both in surface and bulk, for Nitrogen and air-annealed photoanodes, guiding further strategies enhancing the SWS efficiency.

## 2. RESULTS AND DISCUSSIONS

### 2.1. FAST-SWEEP AND STABILIZED PHOTO-VOLTAMMETRY

Photocurrent density, $J_{ph} = J_{ON} - J_{OFF}$, was measured following two protocols that were described in a previous work.[15] For that purpose, we measured the current density for light ON ($J_{ON}$) and respective OFF ($J_{OFF}$), as a function of the applied potential using a three-electrode electrochemical cell, where the photoanode was either the Nitrogen or the air annealed Ti-doped hematite films grown by ACG (S1 and S2, respectively – see sections 4.1 for details about the synthesis). First, voltammetry measurements were performed by varying the voltage between 0.6 and 1.7 V vs. RHE, with a sweep rate of 50 mV/s (Figure S1a). Second, photocurrent density was measured as a function of time while keeping the applied voltage constant. Current density is read once stabilized for an applied voltage in the interval 0.8 – 1.6 V *vs.* RHE. This protocol allows to measure stabilized photocurrent values as the step height between light OFF and ON values (blue and red dot respectively in Figure S1b). The photocurrent values obtained using the two types of measurements (represented with solid lines and symbols for the first and the second approaches, respectively) are represented in Figure 1 for both samples, S1 and S2, as a function of the applied potential. First, we evidence an increase of the photocurrent of more than 200 % at 1.23 V *vs.* RHE for S1 compared to S2 sample. Second, we observe that the photocurrent values measured with both protocols are similar. This point was exhaustively discussed in a previous work[15] and related to low transients that can be observed on the ON – OFF measurements, for annealing temperatures of more than 600 °C.

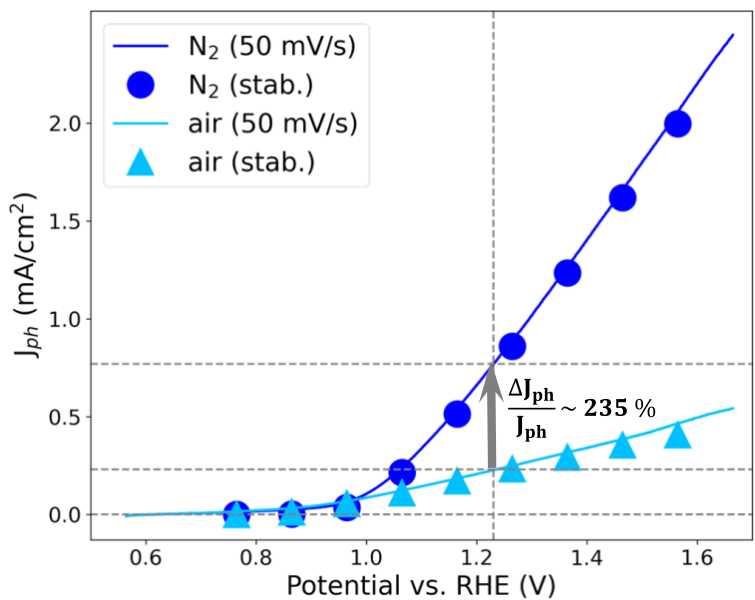

**Figure 1.** Photocurrent as a function of the applied voltage for S1 (dark blue) and S2 (light blue) samples. Photocurrent values obtained from both protocol measurements are reported: fast-sweep rate (solid lines) and stabilized (symbols). A strong photocurrent increase is observed, of more than 200 %, at 1.23 V *vs.* RHE for S1 with respect to S2. At 1.6 V *vs.* RHE, the photocurrent increases by a factor of 4.

2.2. FLAT BAND POTENTIAL AND CARRIER DENSITY

Electrochemical impedance spectroscopy (EIS) allows characterizing the electrical properties of photoanodes, such as the quantification of donors density ($N_d$), the estimation of flat band potential ($V_{fb}$), band bending and charges separation at the interface between the semiconductor and the electrolyte. Nyquist and Mott-Schottky plots are common analytical tools generally used in electrochemistry to model the equivalent circuit that better describes the interface in such systems. Using the Mott-Schottky analysis we can determine $N_d$ and $V_{fb}$ values by a simple linear fit to the experimental data expressed as $1/C^2$ as a function of the applied potential. This analysis works well for model systems: homogeneous and continuous photoanodes, sufficiently thick. In our case, the photoanodes consist of nanorods, with sizes of few hundreds of nanometers, perpendicular to the FTO substrate. The linear fit from the Mott Schottky analysis does not give accurate absolute values for $N_d$ and $V_{fb}$ [33,34] and therefore they should not be simply compared to the literature, especially when applied on photoanodes presenting different morphology. We use these values here only to compare the electric properties between the S1 and S2 samples.

2.2.1. NYQUIST PLOTS

The Nyquist representation plots the imaginary part, $|Z_{img}|$, as a function of the real part, $Z_{re}$. Figure S2 presents the Nyquist plots for the S1 and S2 photoanodes, for several voltages ($V$) between 0.7 V and 1.6 V *vs.* RHE, and for different frequencies ($f$) between 10 Hz and 900 kHz. On each plot we distinguish two regions: a) high frequency region (from 3 to 900 kHz) corresponding to a semicircle characterizing the internal resistance and RC-like semiconductor impedance, and b) low frequency region (from 10 Hz to 3 kHz) —characterizing both the Helmholtz layer formation in the electrolyte near the interface with the photoanode and ion diffusion through the Helmholtz layer. First, we observe reduced impedance values for S1 compared to S2. This result reflects a better overall bulk conduction in the first case, due most probably to oxygen vacancies. Indeed, considering that the Ti doping is similar between the S1 and S2 samples, the conduction difference could be explained by oxygen vacancies produced by the annealing under Nitrogen. Second, the low frequency region exhibit two regimes (slopes) as a function of the applied voltage: i) for V < 1.5 V, charges accumulate in the Helmholtz layer and the capacitive character of the impedance increases (larger $|Z_{img}|$ for similar $Z_{re}$); ii) for V > 1.5 V, we observe a reduced slope value, originating from ion diffusion through the capacitive layer.[15]

2.2.2. DETERMINATION OF FLAT BAND AND CARRIER CONCENTRATION

Experimental $Z(f)$ spectra were fitted using the equivalent circuit $(R1||Q1) - (R2||Q2) - R3$ shown in Figure S3. $R1$ ($R2$) and $Q1$ ($Q2$) are the resistances ($Ri$) and the constant-phase-elements (CPE, $Qi$) associated to the semiconductor photoanode (Helmholtz layer). As detailed in Supporting Information, the values of the capacitance, $C$, associated with the space charge layer in the semiconducting photoanode in contact with the electrolyte, can be calculated with Equation S4, using $Qi$ and $N_i$ parameters extracted from the fit (Figure S4). Mott-Schottky plots, expressing $1/C^2$ as a function of the voltage obtained at 900 kHz for both photoanodes, are presented in Figure 2. It is important to mention here that Mott-Schottky plots obtained for all frequencies in the interval 10 Hz – 900 kHz are similar. $C(f)$ slopes for multiple potential values are quasi-parallel, particularly for applied potentials between 1.1 and 1.4 V (Figure S5). The simplest model used to fit Mott-Schottky plots supposes a linear variation with the voltage of $1/C^2$. In reality, this linearity depends on several parameters: the semiconductor state (accumulation, depletion or inversion), the presence of surface states, the size of the surface features on the photoanode, pH value of the electrolyte, donor levels in the semiconductor, crystallographic structure of the material, etc.[33-37] For a comparative analysis between the S1 and S2 samples we consider that for potentials between 1.1 and 1.4 V (interval materialized by gray region Figure 2), the semiconductor is in depletion state. Therefore, the Mott-Schottky plot can be fitted using the simplified formula:

$$\frac{1}{C^2} = \frac{2}{eN_d\varepsilon_0\varepsilon_r S^2}\left(V - V_{fb} - \frac{K_B T}{e}\right)$$

where $C$ is the capacitance, $e = 1.6 \times 10^{-19}$ C is the electron charge, $N_d$ the carrier concentration, $\varepsilon_0 = 8.85 \times 10^{-12}$ F·m$^{-1}$ and $\varepsilon_r$ are the vacuum and relative permittivity, respectively. Furthermore, $S$ is the sample surface in contact with the electrolyte ($S = 0.5$ cm$^2$ in our case), $V_{fb}$ the flat band, $V$ the applied voltage, $K_B = 1.38 \times 10^{-23}$ J·K$^{-1}$ is the Boltzmann constant, and $T$ is the temperature. A wide range of relative permittivity values for hematite are reported in the literature, depending on the morphology (nanostructures or not), annealing temperature, frequency, etc.[33,38] This results in a significant difference of the predicted capacitance values. In this work, the value used for the relative permittivity is $\varepsilon_r = 10$. Thus, it will be easy for the reader to recalculate capacitance values if one considers different $\varepsilon_r$ values. From the linear fits in Figure 2, we estimate the flat band values, $V_{fb}$, of 0.1 V and 0.8 V and carrier concentrations, $N_d$, of 4.2×10$^{21}$ and 5.4×10$^{20}$ for S1 and S2, respectively. We recall that these values are obtained using a simple linear fit analysis without any consideration about nanorods dimensions (width, length) or density (as number of nanorods / cm$^2$). However, their relative variation is relevant here to compare the electric properties of S1 and S2 samples.

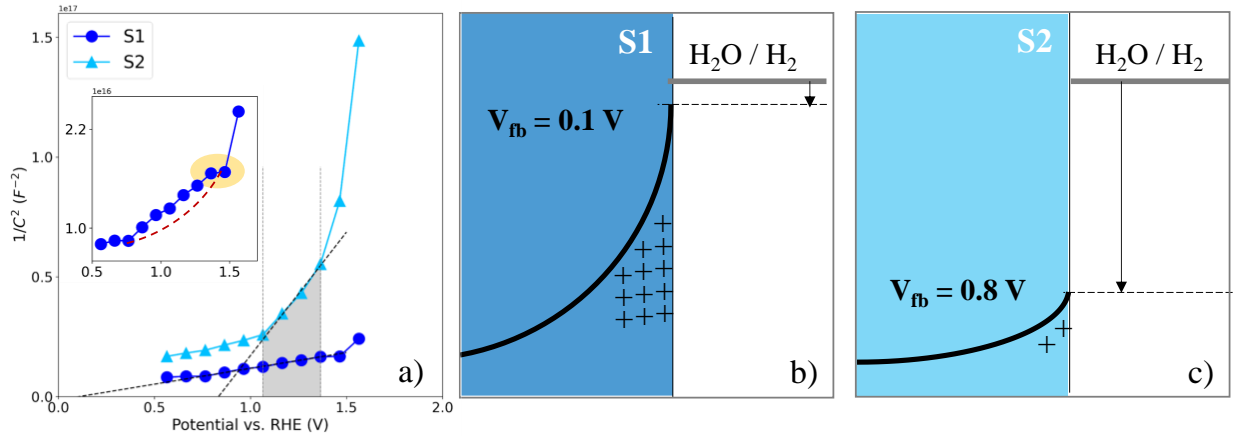

**Figure 2.** Mott-Schottky analysis for S1 (●) and S2 (▲) samples (a). Linear fits were performed for potential values between 1.1 V and 1.4 V vs. RHE, emphasized by the grey region. Larger carrier concentrations values ($N_d$ = 4.2×10$^{21}$) are determined for S1 compared to S2 ($N_d$ = 5.4×10$^{20}$). A yellow region in the inset of (a) indicates the kink region, characteristic to the presence of surface states; b) and c) representation of the band bending at the interface with the electrolyte considering the flat band values estimated for S1 (0.1 V) and S2 (0.8 V) samples, respectively.

It is known that the presence of surfaces states at the semiconductor/electrolyte interface results in the Fermi level pinning, and, in this case, the Mott-Schottky plot presents a plateau (or double kink) where the $C(V)$ is (quasi-)constant.[39,40] In this region, the applied voltage drops within the Helmholtz layer instead the depletion region in the semiconductor.[41] This flat region can be observed indeed in the case of the S1 sample for voltages between 1.4 and 1.5 V (marked region in Figure 2a). To the contrary, nothing is visible for the S2 photoanode. As shown in previous studies[15,42] two kind of surface states may co-exist at the interface between the hematite-based photoanode and the electrolyte, one at higher oxidative energies, lying at the bottom of the conduction band and a second at lower energies. While the first kind is beneficial for the PEC process, the second plays a detrimental role as being inactive from the PEC point of view, driving recombination of charges at the interface. In the present case, we identify the first kind (i.e. beneficial, at higher oxidative energy) of surface states leading to lower flat band value for the S1 sample. This determines higher band bending at the interface with the electrolyte (Figure 2b), and, consequently, better charge separation. Furthermore, we find a carrier concentration ten times higher for S1 than for S2, which confirms previous observations related to improved electric conduction. Therefore, we can conclude that there are two mechanisms determining higher photocurrent induced by Nitrogen atmosphere annealing: i) at surface, due to better charges separation promoted by surface states, and ii) in bulk, by improved electric conductivity due to higher carrier concentration, related to oxygen vacancies.

It is interesting to focus here briefly on the preponderance of each of these mechanisms. First, it is an accepted fact that Ti doping increases the electrical conductivity acting as *n*-type doping in the hematite semiconductor, increasing thus the measured photocurrent during the PEC process. Figure S6 presents the comparison between two additional samples, one annealed in air and the other under Nitrogen, containing 30 times less Ti compared to S1 and S2 samples. On one hand, it is important to note that even at this very low Ti content level the photocurrent is increased compared to bare hematite. There is barely any visible difference between the two cases, the sample annealed under Nitrogen exhibiting only slightly higher photocurrents above 1.3 V vs. RHE. It appears thus that for very low Ti contents, the Nitrogen annealing has marginal effects on the recorded photocurrent. We can conclude therefore that the increase expected from additional oxygen vacancies doping, formed upon annealing under Nitrogen, is negligible. Consequently, the photocurrent enhancement we determine for the S1 sample compared to the S2, is mainly due to the formation of surface states promoting enhanced charge separation.

## 2.3. SURFACE *VS.* BULK CHEMICAL COMPOSITION AND COORDINATION

### 2.3.1. Shadow S-XPEEM

We combine the photoemission surface sensitivity (electron mean free path of few nanometers) with the capability of S-XPEEM to discriminate surface and bulk information (for details see Supporting Information, section III). To this purpose, we measured energy stacks by recording full-field XPEEM images while tuning the X-rays energy across the Ti $L_{2,3}$, O K and Fe $L_{2,3}$ absorption edges, from 450 to 750 eV. Figure 3 shows the results obtained for S1 and S2 samples. The RGB color-coded images (Figure 3a for S2 and 3b for S1, respectively) are obtained using the singular value decomposition (SVD) method, applied after correcting the drift of the hyperspectral data, using the SVD implementation in the aXis2000 software.[43] We identify three main spectral contributions, each one represented by an associated color in Figure 3: bulk one, recorded from the shadow region (Red), gold-plated Si substrate (Green), and surface contribution (Blue) (see also Figure S7 for further details).

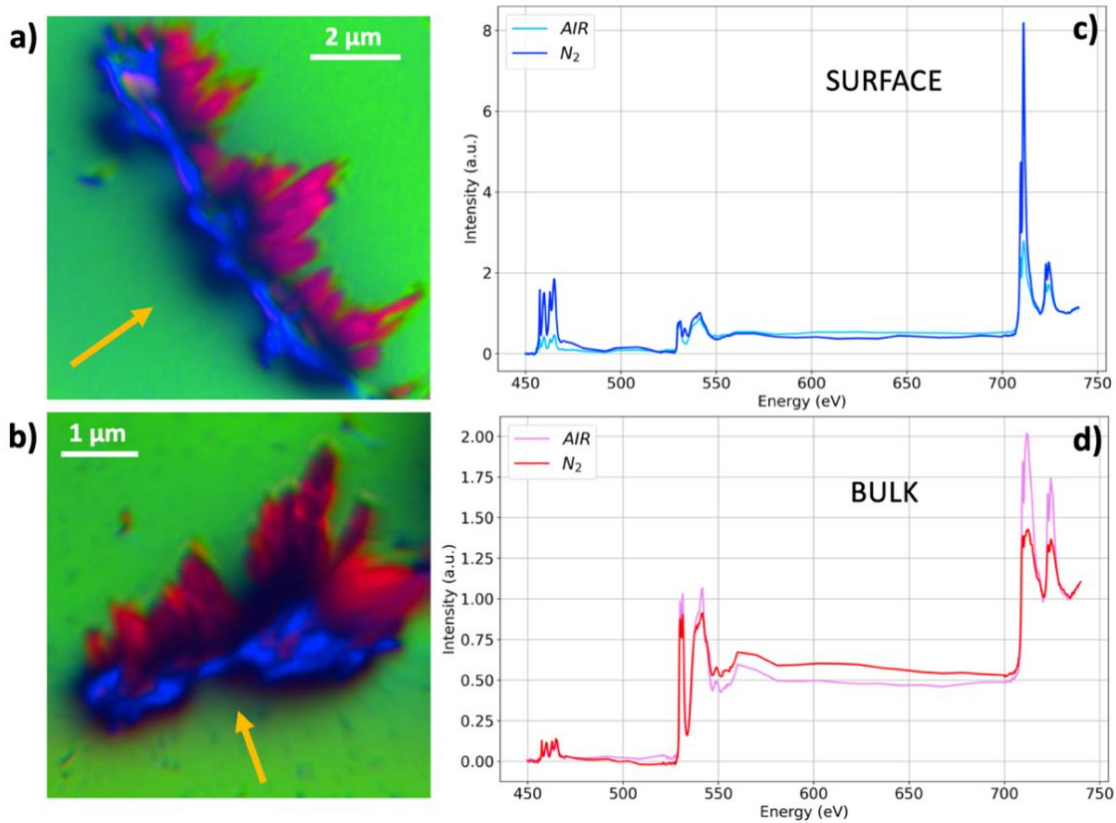

**Figure 3.** S-XPEEM results from the $N_2$ (S1) and air annealed (S2) samples; a) and b) RGB color-coded map representations of the 3 regions of interest: **R**ed – bulk (shadow), **G**reen – gold substrate, **B**lue – surface, respectively; the orange arrow indicates the direction of the X-rays beam; the images are kept in the original orientation as given by the electron microscope lens settings for the respective field of views used: 20 µm for S2 samples in a) and 10 µm for S1 sample in b); c) comparison of the surface signals recorded from the corresponding blue regions of both samples: we remark the strong spectral differences at the Ti $L_{2,3}$ (450 – 470 eV) edges and O K edge (520 – 560 eV); d) comparison of the bulk signal recorded from the corresponding red (shadow) regions of both samples: the spectra look very similar, signature of a similar bulk chemical nature of the two samples.

The images shown in Figure 3a and 3b are kept in their original orientation, related to the use of 20 µm and 10 µm fields of view, respectively. Indeed, in the XPEEM microscope column, the image is turned by the specific lens settings for each field of view, while the detection part is kept in fixed orientation. Thus, to orient the reader, the direction of the X-ray beam is materialized using the orange arrow. Surface and bulk (shadow) regions can be easily distinguished, marked as blue and red regions, respectively. The size of the Ti-doped hematite particles deposited on the gold-plated substrate was estimated from their shadow size: it spans from 200 to 500 nm, taking into account the 16 degrees incidence angle of the impinging X-rays with respect to the sample surface. To compare the chemical composition between the samples, *i.e.* the elemental Ti/O/Fe ratio, the spectra are normalized in two steps. First, a normalization is applied with respect to the incoming X-ray beam intensity, $I_0$ (Figure S8). To fulfil the requirement of a flat spectral response over the useful energy range, gold-plated substrates were used to obtain the $I_0$. Since we record two kinds of signals on the sample, *i.e.* photoemission (surface – blue) and transmission (bulk – red), different $I_0$ normalization are required. Thus, for the surface signal, we used $I/I_0$ (Figure S8c). The optical density of the bulk signal was extracted as $OD = log(I_0/I)$ (Figure S8b). A second normalization was applied, considering that the chemical formula includes only Ti, O and Fe. A linear background is subtracted from the Ti $L_{2,3}$ pre-edge region (450 – 455 eV) bringing this region to zero. In this way we remove spectral contribution arising from thickness and density. Then the spectra are matched over the Fe $L_{2,3}$ post-edge (735 – 740 eV) region, by applying a multiplicative factor. As a result, the elemental chemical ratio is directly proportional to the edge jumps. As it can be observed in Figure 3d, the spectra corresponding to the bulk regions are very similar for the S1 and S2 samples. Indeed, the edge jumps are almost identical suggesting that there is no difference in the bulk elemental composition between S1 and S2. Across the Fe $L_{2,3}$ absorption edges, the signal from bulk is saturated due to the thickness of the particles (*i.e.* larger than 200 nm) and the strong X-ray attenuation length that varies between 80 and 120 nm in this spectral region. There are no differences at the resonant energies of Ti, O and Fe. Contrary to the spectra from the bulk regions, the surface ones show higher edge-jump values across the Ti $L_{2,3}$ absorption edge for both samples (Figure S9). This suggests surface Ti-segregation, as we already shown in a previous study dedicated to the effect of Ti doping of hematite photoanodes.[15] The S1 sample exhibits even higher edge jump (Figure 3c) than S2, related to a stronger Ti surface segregation. Thus, we can conclude that the annealing under Nitrogen atmosphere promotes Ti segregation toward the surface of the Ti-doped hematite photoanodes.

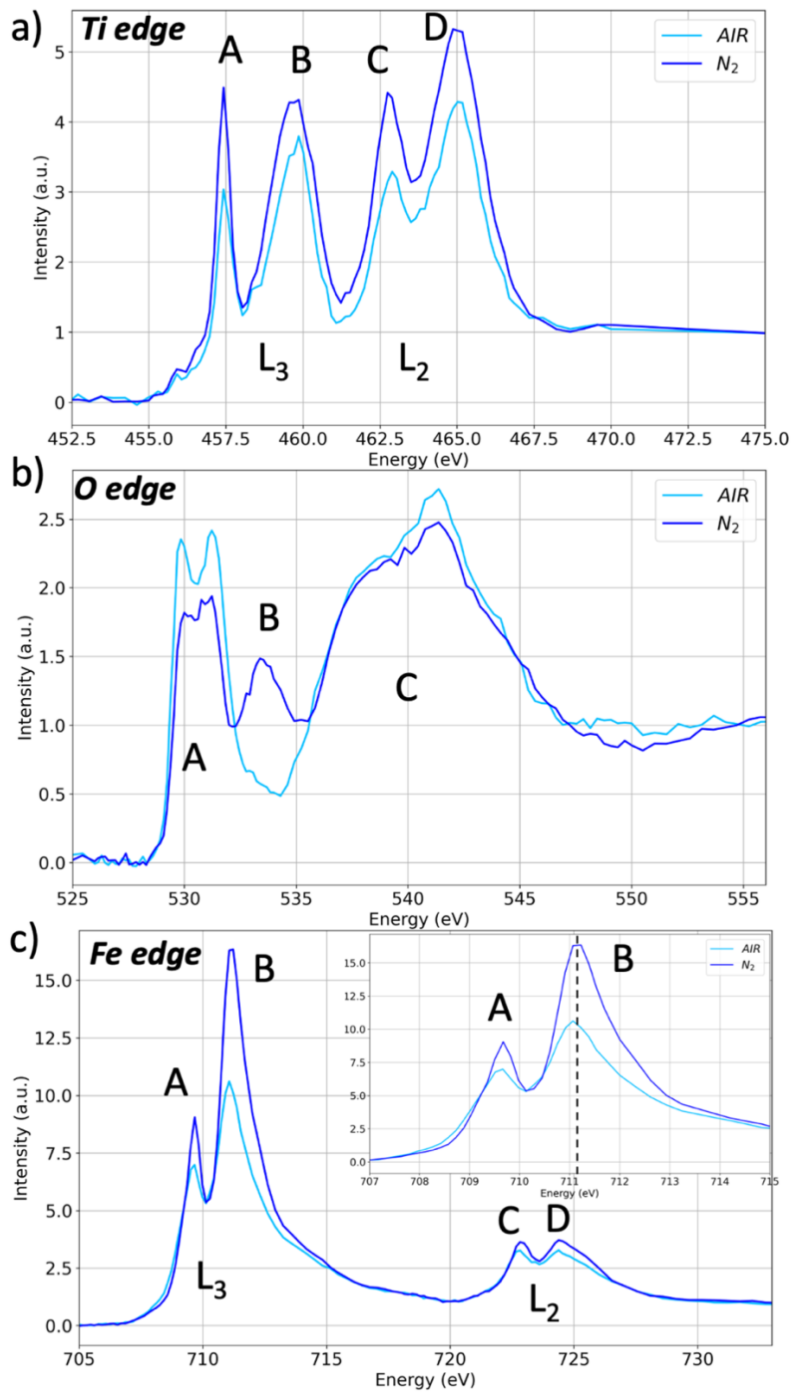

**Figure 4.** NEXAFS spectra extracted from the surface regions of the S-XPEEM measurements across the Ti $L_{2,3}$ (a), O K (b), and Fe $L_{2,3}$ (c) X-ray absorption edges. Spectra are normalized at each of the absorption edges such as to address only the chemical coordination of the probed ion. The major spectral features are labeled as follows: a) Titanium: A – $L_3$ $t_{2g}$, B – $L_3$ $e_g$, C – $L_2$ $t_{2g}$, D – $L_2$ $e_g$; b) Oxygen: A – hybridized spectral region showing Fe $t_{2g}$ – $e_g$ and Ti $t_{2g}$ crystal field splitting, B – hybridized Ti $e_g$, C – hybridized Fe and Ti 4sp bands; c) Iron: same notations as for Titanium in a).

Near-Edge X-ray Absorption Fine Structure (NEXAFS) spectroscopy can discriminate fine chemical coordination particularities. Spectral features in the resonant region of the X-ray absorption edges are related to the electronic structure of the probed ion. Figure 4 presents three spectral regions of interest. With respect to the Figure 3c, where we show spectra normalized over the whole energy range (450 – 740 eV), here we apply similar normalization, *i.e.* we subtract linear background in the pre-edge region and multiply the spectrum to match the post-edge, but limited to each elemental absorption region independently. We can thus focus on the electron structure comparison between samples. At Ti absorption edge, Figure 4a, the S2 sample presents a spectral shape characterized by symmetrical B (Ti L$_3$ e$_g$ electronic transition) peak, likewise in ilmenite (FeTiO$_3$) structure (Figure S10).[44] The relative intensities (branching ratio) between the A, B, C and D peaks, are equally characteristic to the Ti$^{4+}$ ion as expected for an ideal Ti substitution of the octahedrally coordinated Fe$^{3+}$ ion in the hematite (Fe$_2$O$_3$).[15] Is important to note here that we do not measure ilmenite, since not getting the needed stoichiometry, but it can serve as surrogate to describe the local structure of the substitutional Ti-doping of the hematite structure. This local structure is probed by NEXAFS and is thus responsible for the specific spectral features. For the S1 sample the A:B branching ratio changes, the Ti L$_3$ t$_{2g}$ band presents increased intensity compared to S2, while the C:D branching ratio is the same. The spectral shape is similar to TiO$_2$ brookite structure.[44,45] Considering the chemical information detailed above, we cannot consider TiO$_2$ alone at the surface of S1 sample. It is more reasonable to consider the pseudo-brookite structure, Fe$_{1+x}$Ti$_{2-x}$O$_5$, with *x* between 0 and 1. The corresponding crystallographic structures are represented in Figure S10. From the spectral shape it is impossible to obtain a particular value for *x*, since the Ti$^{4+}$ ion has the same octahedral coordination (Ti-O$_6$) over the entire x range, from 0 to 1. Additional information can be extracted from the NEXAFS Fe spectra in Figure 4c. First, we remark that, contrary to what one would expect, the S2 sample presents slight reduction of the Fe$^{3+}$, the B spectral line is shifted toward lower binding energies by < 0.1 eV with respect to the S1 sample. This leads also to a variation of the *d*-orbital splitting, $t_{2g} - e_g$ (A-B), from 1.48 eV for S1 to 1.41 eV for S2. Such behavior is related to distortions of the Fe-O$_6$ octahedra very close to the hematite nanoparticles surface and the Ti substitution in the hematite structure.[46] Since S1 sample presents increased Ti content toward the surface, this distortion is further accentuated. In addition to the structural origin of this behavior, we have to consider that in the presence of Ti substitution, part of the Fe$^{3+}$ from the hematite structure is replaced by Ti$^{4+}$ and iron is reduced (Fe$^{2+}$), following the charge conservation relation: $2 \cdot Fe^{3+} = Fe^{2+} + Ti^{4+}$. Therefore, for the S2 sample, the slight iron reduction agrees with the spectral shape measured at Ti absorption edge (Figure 4a) and is related to formation of Ti-O$_6$ octahedra upon Ti substitution of Fe. Conversely, S1 sample exhibits increased proportion of Fe$^{3+}$, signature for the formation of the pseudo-brookite phase, Fe$_{1+x}$Ti$_{2-x}$O$_5$ with $x$ = 1,[47] capable of holding high concentration of Fe$^{3+}$ oxidation state from charge equilibrium criteria, in agreement with the NEXAFS spectra recorded at Ti edge (Figure 4a). Unfortunately, the literature lacks extended reports of pseudo-brookite NEXAFS at the Fe L$_{2,3}$ absorption edge for further comparison. Lv et al.[31] reported Fe L$_{2,3}$ NEXAFS measurements recorded in total electron and fluorescence yield modes, for either coating or incorporated Fe$_2$TiO$_5$ into hematite. They evidenced only Fe$^{3+}$ related to the formation of the pseudo-brookite phase. A second notable feature that can be extracted from the Fe L$_{2,3}$ NEXAFS, is the variation of A:B branching ratio that has a value of 0.65, for S2, that is expected for hematite.[46] To the contrary, S1 exhibits a value of 0.56, equivalent to an iron oxyhydroxide (e.g. goethite). The branching ratio is related in first approximation to the Fe oxidation state. The Fe$^{3+}$ is partially reduced in S2, as discussed above. This could explain alone the branching ratio variation between the two samples. However, we cannot exclude other factors that might be involved, like the ligand field strength (or crystal field strength), that varies from 1.48 to 1.41 eV.

The O K-edge NEXAFS (Figure 4b) is of particular interest for oxide materials in general, due to the delocalized character of the O 2p orbitals. These orbitals are responsible thus for their conduction properties.[48] We probe in this case the unoccupied electronic states available in the conduction band. The spectra exhibit two distinct regions, below and above ~535 eV. In the region marked as A, we probe electronic transitions from O 1s to hybridized Fe 3d – O 2p bands, showing d-orbital splitting, $t_{2g}$ and $e_g$.. In the region marked as C, there are mixed O 2p and hybridized Fe 4sp – O 2p bands. The S2 sample shows typical features related to the Ti-doped hematite, as already shown:[15] prominent Fe 3d – O 2p hybridization, and the presence of oxidized surface states (OSS) marked by a strong asymmetry in the spectral region labelled with B. At the same energy position, the S1 sample displays an additional peak. The d-orbital splitting value for S1 is clearly reduced compared to S2, in agreement with the information extracted from the spectra recorded at Fe $L_{2,3}$ edge (Figure 4c). Owing to the $Fe_2TiO_5$ formation at the surface of the sample, we probe a mixture of hematite and titanate character. Before any comparison with the literature, it is mandatory to note that NEXAFS cannot give absolute energy positions unless precise X-rays energy calibration is performed. That is not always the case. Therefore, reported values may vary slightly. Moreover, it is useful to compare energy positions between hematite and titanates data recorded using the same experimental conditions, as in F.M.F. de Groot et al.[49] In that report, similar to our finding, the O 1s – Fe $e_g$ and O 1s – Ti $t_{2g}$ bands are clearly degenerate. Hereafter we express all energy positions relative to O 1s – Fe $t_{2g}$, at 529.9 eV, and then consider relative positions from the literature.

Table 1: absolute and relative Fe and Ti peak positions in eV for the S1 sample, with respect to the Fe t2g one (first column), extracted from the O K-edge NEXAFS spectra and compared to relevant data from literature. OSS position is identified at +2.2 eV from Fe t2g. † expected position of the O 1s – Ti eg considering +2.7 eV with respect to the O 1s – Fe t2g.

| Fe $t_{2g}$ (eV) | Fe $e_g$ / Ti $t_{2g}$ (D) (eV) | OSS (D) (eV) | Ti $e_g$ (D) (eV) | reference |
|---|---|---|---|---|
| 529.9 | 531.3 (+1.4) | 533.5 (+2.2) | 534.0 (+2.7)† | this work |
| 529.4 | 530.7 (+1.3) | - | 533.3 (+2.6) | 49 |
| - | 530.8 | - | 533.5 (+2.7) | 50 |

Table 1 summarize the positions of hybridized Fe and Ti bands from the O K-edge NEXAFS. We report in parenthesis the relative positions (D) with respect to O 1s – Fe $t_{2g}$ at 529.9 eV. The values extracted from Figure 4b are shown in the first row. The 4$^{th}$ column indicates the expected position for the O 1s – Ti $e_g$ transition, at D = +2.7 eV with respect to the O 1s – Fe $t_{2g}$ at 529.9 eV.[50] Typically, Fe d-orbital splitting in hematite is 1.4 eV, while Ti d-orbital splitting for $Ti^{4+}$ ion, like $TiO_2$, is 2.7 eV. The peak position we could assign to O 1s – Ti $e_g$ is 533.5 eV, *i.e.* D = +2.2 eV instead of the expected one at D = +2.7 eV. To explain this difference, one could consider strong distortion effects of the octahedral symmetry, or a highly disordered surface,[51] affecting the expected crystal field splitting. But in this case, the dip at ~532 eV between the A and B spectral regions in Figure 4b, would be less pronounced compared to our spectra. Thus, we cannot assign the B peak in Figure 4b to pure contribution of Ti $e_g$ band and we address this peak as signature of OSS. The presence of these surface states, with highly oxidized character, leads to increase of charge transfer rate between the photoanode and the electrolyte, increasing the overall PEC activity.[15] We probe the same surface states as the ones suggested by the PEC data. Indeed, photocurrent measurements directly probe the contribution of these OSS at the overall charge transfer, while NEXAFS probes OSS as unoccupied levels through transition of core level electrons.

Is important to mention that we did not find any Sn diffusion, as often reported for the hematite based photoanodes deposited on FTO substrates.[52] At the O K-edge (~533 eV) we would expect a characteristic spectral feature related to electronic transitions in the Sn 5p orbitals. This is not the case for any of the samples studied here. It is generally accepted that Sn diffusion scales with the annealing temperature. Since both samples were annealed at the same temperature (*i.e.* 600 °C), if any Sn diffusion is present (not detectable by our measurements), it should be of the same amount and therefore affect the PEC results at same level. In addition, we tested the possible formation of nitride phases at the N K-edge (for Fe-N and/or Ti-N coordination). Nothing detectable was measured. We want to stress that, as for the Sn diffusion, NEXAFS loses its sensitivity for concentration values below ~0.1 %. While we cannot rule out definitively the formation of nitride phases, that would be certainly below the needed quantity to drive the PEC enhancement upon $N_2$ annealing. In addition, we recall that the probability of obtaining nitrides by exposing an oxide in $N_2$ atmosphere, even at higher temperatures, is very low (related to the high activation energy barrier of the N-N triple bond of 940.95 kJ/mol). Indeed, the studies reporting metal nitrides synthesis show different methods than high temperature annealing.[53]

2.3.2. CORRELATIVE MEASUREMENTS: STXM, SPECTRO-PTYCHOGRAHY AND STEM

S-XPEEM offers high surface selectivity, but its spatial resolution is altered due to space charge effects,[54] occurring particularly in 3D-shaped samples, like the ones measured here. Space charge effect arises from Coulomb collective repulsion of the photoelectrons generated in the surface region of our samples, resulting in a spatial broadening and therefore in resolution loss. To improve the spatial resolution associated with the spectral features discussed above, we measured the samples using STXM and spectro-ptychography. These are photon-in/photon-out techniques, exempted thus from space charge effects. Ptychography allows accessing improved resolution compared to conventional STXM, through ptychographic reconstruction of series of X-rays scattering images originating from overlapping regions on the sample.[55,56] This is, to the best of our knowledge, the first report of spectro-ptychography at Ti $L_{2,3}$ absorption edge, enabled by the use of an sCMOS camera equipped with sensor providing high efficiency at low energies.[57,58] Further details about the ptychography implementation used here are given in the Methods section. All absorption edges (Ti $L_{2,3}$, N K, O K, Fe $L_{2,3}$) were measured using STXM (*i.e.* using a photomultiplier tube detector) to identify first spectral features of interest on both S1 and S2 samples. It is possible thus to focus on specific spectral range to perform spectro-ptychography measurements, a time-consuming approach both for acquisition and reconstruction. Thus, Ti $L_{2,3}$ edge was found most adapted for the correlative description detailed in the following. Indeed, it presents outstanding spectral features related to the chemical nature of the pseudo-brookite from the surface region, probed also by S-XPEEM. The 3D spectro-ptychography dataset included as Supplementary Information, was measured additionally across the O K-edge where we observe same OSS spectral position as in S-XPEEM (*i.e.* 533.5 eV).

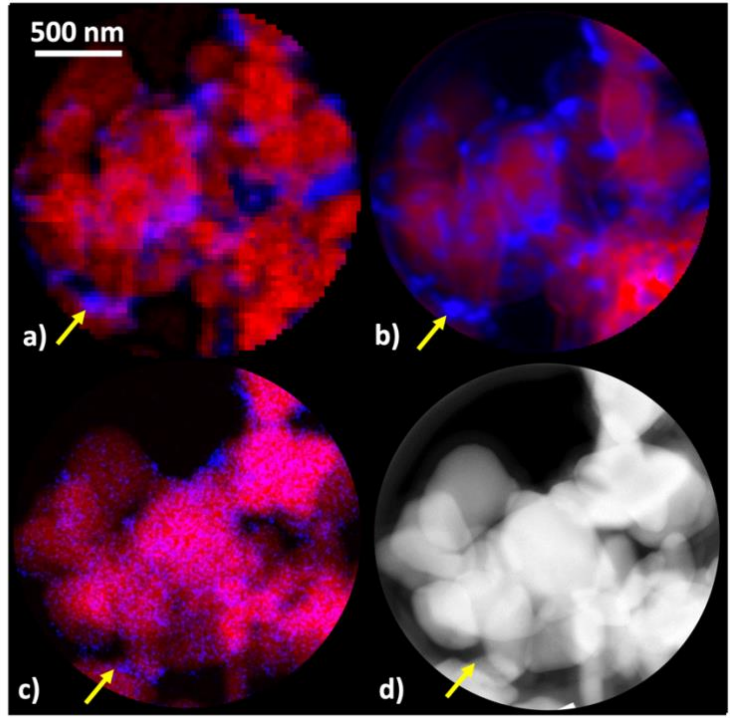
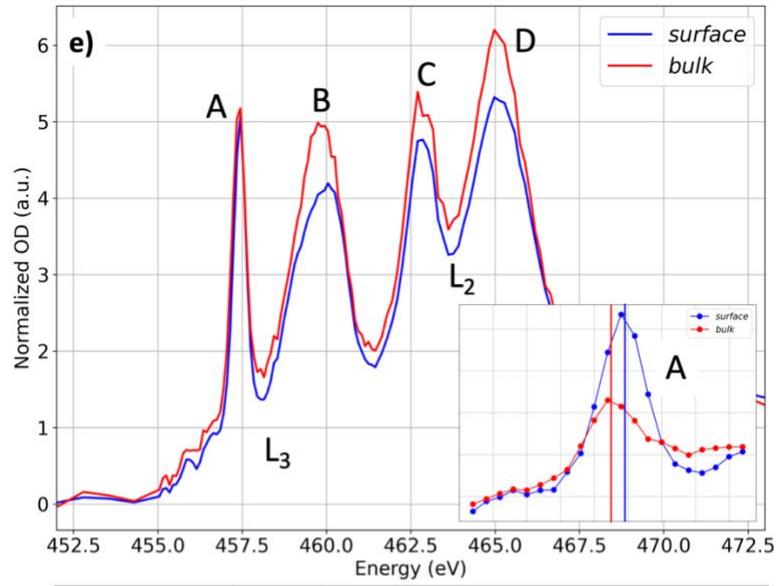
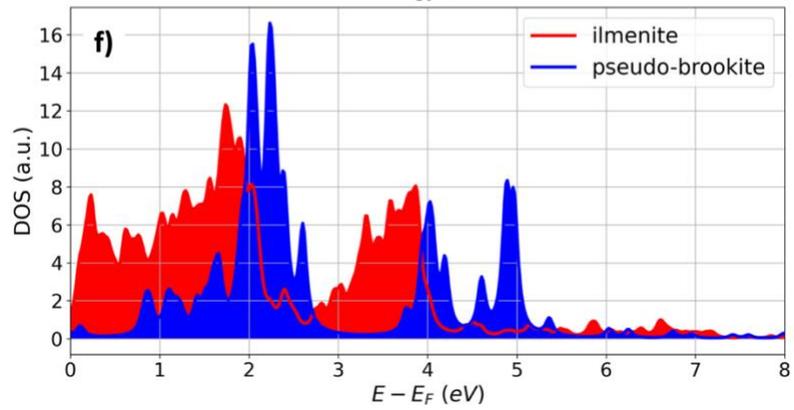

**Figure 5.** Overview of STXM, spectro-ptychography, and STEM (EDXS and HAADF) results obtained from the same S1 sample region, along with DFT calculations. Speciation maps obtained from STXM hyperspectral data (a), spectro-ptychography (b), elemental mapping from EDXS (c), and the corresponding HAADF image (d). Circular overlays were used in the images matching the support hole size (2 µm) to keep only the signal from the unsupported sample part (details in text and Supporting Information). The yellow arrows are guide to eyes pointing to the same particular position on the sample; e) Ti $L_{2,3}$ edge NEXAFS spectra of the surface $Fe_2TiO_5$ clusters (blue) and bulk Ti-doped hematite (red), used to obtain the speciation maps in a). The inset showing the spectral region (A) Ti $L_3$ $t_{2g}$, used for spectro-ptychography speciation map (b), exhibiting +0.1 eV energy shift for the surface component; f) summed $t_{2g}$ and $e_g$ bands calculated by DFT using model bulk structures for ilmenite (red) and pseudo-brookite (blue). Pseudo-brookite exhibits energy shift toward higher binding energies, strongly increased $t_{2g}$ and slightly reduced $e_g$ integrated DOS.

Figure 5 presents the results obtained from the S1 sample using STXM, spectro-ptychography and STEM. The same nanoparticles were measured using STXM (Figure 5a), spectro-ptychography (Figure 5b) and STEM in EDXS (Energy Dispersive X-ray Spectrometry) and HAADF (High Angle Annular Dark Field) modes (Figure 5 c and 5d, respectively). This was possible using special hole-tagged SiN membranes (see Supporting Information – IV for details) which can be consecutively employed by different microscopy techniques. Yellow arrows point to the same particular sample position as guide to the eyes. Using STXM, we measure full NEXAFS spectral region across Ti $L_{2,3}$ absorption edge, with high spectral resolution, in this case better than 0.1 eV. Indeed, hyperspectral data can be obtained with large number of spectral points, while keeping a reasonable spatial resolution, adapting the scanning step-size at the beam spot-size of 30 nm. In this manner we isolated two specific spectral components (Figure 5e), marked with blue and red colors, assigned for surface and bulk signals, respectively. The corresponding R(G)B color-coded map presented in Figure 5a was obtained using the SVD method implemented in the aXis2000 software after applying a drift correction to the recorded stack, similarly to the S-XPEEM datasets (Figure 3a and 3b). We applied a circular overlay of 2 µm diameter corresponding to the 2 µm diameter holes in the SiN support membrane. Thus, the $I_0$ signal was taken from the free space around the nanoparticles. Surface clusters (blue) and bulk nanoparticles (red) signals were isolated correspondingly (Figure 5e). They are similar to those obtained from the S-XPEEM (Figure 4a): increased A:B branching ratio for the surface component of S1 sample. NEXAFS spectra obtained by STXM exhibit two major additional features. First, Ti $L_3$ $t_{2g}$ peak corresponding to the surface clusters (blue) is shifted by +0.1 eV. Second, the surface component (Figure 5e) presents an asymmetric B peak, that could not be evidenced using the S-XPEEM, because of the poorer spatial definition in that case. This asymmetry agrees with the pseudo-brookite character of the surface clusters,[44] as suggested by S-XPEEM. The spectral shape of the bulk component (red in Figure 5e) confirms substitutional $Ti^{4+}$ inside the hematite structure.[59]

Switching to spectro-ptychography is very handy at HERMES beamline. The standard 0D detector, a photomultiplier tube, is replaced with a 2D detector (sCMOS camera), without modifying the sample setup (see Methods section 4.4.3. and Supporting Information – IV). Spectro-ptychography data was recorded from the same region of interest (ROI) as the one used for STXM (Figure 5b). The R(G)B color-coded map was obtained measuring 21 energy points across the Ti $L_3$ $t_{2g}$ (peak A in Figure 5e and inset) transition, using SVD similarly to S-XPEEM and STXM. Compared to STXM, less spectral points can be measured due to the huge amount of generated data (*e.g.* more than 14 Go per X-rays energy point), difficult to analyze even using dedicated workstation equipped with multiple GPUs. We can estimate that surface clusters (blue) size span from ~10 nm to ~100 nm, considering the ptychographic reconstructed image pixel size of 8.5 nm (Figure 5b). We mention that STXM spectral components shown in Figure 5e were extracted as a second iteration during the cross analysis between STXM and spectro-ptychography data. Indeed, spectro-ptychography allows to isolate a spatial region with pure "blue" component, associated to the surface clusters. Owing to NEXAFS strong chemical coordination sensitivity we can separate distinct spectral components arising from distinct spatial locations. As a result, the chemical coordination allows improving spatial resolution using spectro-ptychography (see also Figure S12). Spectro-ptychography recovers both amplitude (absorption) and phase (dispersion) information (Figure S12). Further analysis can be performed following the arguments used by Farmand et al.[60] Such in-deep analysis of spectro-ptychography is however beyond the scope of this paper. Analytical STEM-EDX (Figure 5c) characterization performed on the same nanoparticles, confirms the presence of the Ti-rich area at the surface of the Ti-doped hematite particles. The HAADF image recorded simultaneously with the EDX one, is also shown (Figure 5d).

We performed DFT calculations to confirm the NEXAFS spectral features at Ti $L_{2,3}$ absorption edge, *i.e.* +0.1 eV energy shift of the $L_3$ $t_{2g}$ and the variation of A:B branching ratio. For this purpose, we used ilmenite bulk structure, as model for Ti-doped hematite, and pseudo-brookite structure, as shown in Figure S10. First, it is important to remind that NEXAFS probes unoccupied states via electronic transitions from core levels. DFT allows calculating the contribution to the density of states (DOS) of each orbital per atom of the unit cell. To compare DFT calculations with NEXAFS experimental data, Figure 5f shows the summed contributions of the Ti 3d orbitals for ilmenite (red) and pseudo-brookite (blue), respectively. The $t_{2g}$ level (lower binding energy) is composed from $d_{xy}$, $d_{xz}$ and $d_{yz}$ orbitals, while $e_g$ (higher binding energy) gathers $d_{x^2-y^2}$ and $d_{z^2}$ ones. We performed the synchrotron X-rays measurements using circular polarization to remove any contributions arising from charge anisotropies in the sample. Thus, if the pseudo-brookite clusters are isotropically oriented, summing over all orbital orientation is valid for a qualitative comparison with the experimental data. Calculated DOS reproduce the experimental NEXAFS spectra features. First, pseudo-brookite exhibits an energy shift toward higher energies. The calculated shift value (~ 1 eV) is much larger compared to the experimental measured value (0.1 eV). This is probably related to the approximation made here, *i.e.* considering ilmenite structure to model the Ti-doped hematite. Indeed, the estimated Ti substitution level for our samples is less than 10 %, while ilmenite structure can be assimilated to 50 % Ti doping. Second, the experimental variation of the branching ratio A:B (Figure 5e) is reproduced by the calculated DOS. Thus, the pseudo-brookite has increased A:B ratio compared to ilmenite. In addition, pseudo-brookite DOS presents two distinct features at +4 eV and +5 eV relative to the Femi level, that can be associated to the asymmetry of the $e_g$ measured by NEXAFS (B peak in Figure 5e).

## 2.4. CRYSTALLOGRAPHIC STRUCTURE

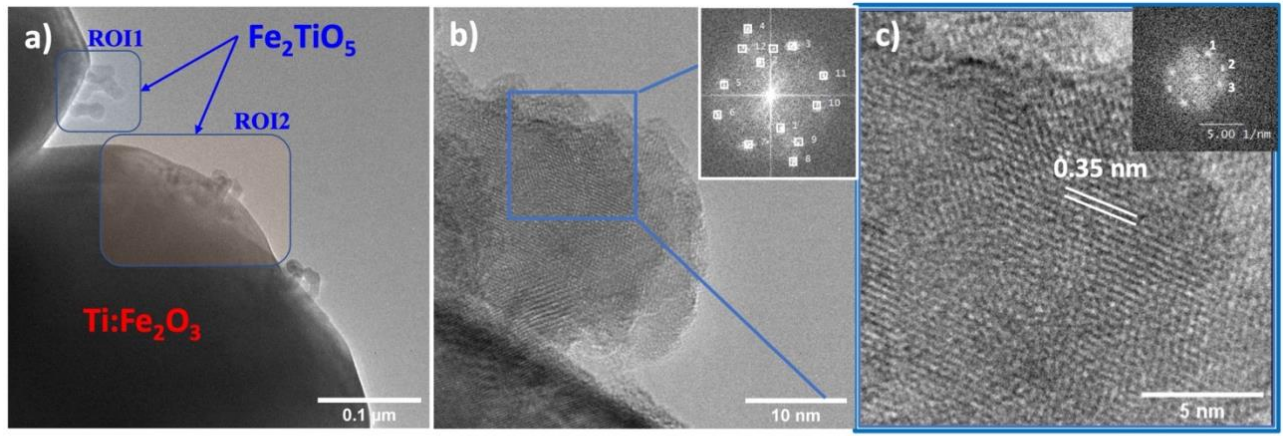

Figure 6. STEM crystallographic measurements of S1 sample: a) image showing Ti-doped nanoparticle with pseudo-brookite surface clusters. Two distinct ROI are indicated presenting two types of morphology for the pseudo-brookite clusters; b) high resolution STEM cliché with associated FFT (Fast Fourier Transform) in the insets; c) cluster zoomed image showing well defined atomic planes with the associated FFT inset presenting bright diffraction spots corresponding to the pseudo-brookite structure; d) electron diffraction indexation with the 3 main spots arising from the pseudo-brookite structure.

Structural analysis performed by high resolution STEM (Figure 6) consolidates the spectro-microscopy results. Figure 6a shows the detail of a Ti-doped hematite nanoparticle with pseudo-brookite surface clusters. The size of these clusters, of few tens of nanometers, fits well the average size determined from spectro-ptychography. We observe two types of morphology for the pseudo-brookite clusters: one with pronounced 3D character (columnar-like) (Figure 6a – ROI1), with long chains composed of few clusters and where the interface between hematite nanoparticles and pseudo-brookite clusters is quite small (see also Figure S13). The second one is characterized by clusters lying on the surface of the hematite nanoparticles (Figure 6a – ROI2). Both morphologies can be recognized in Figure 5b obtained by spectro-ptychography, where we can identify isolated and elongated "blue" spots exhibiting pseudo-brookite spectral features. The similarity and complementarity between STEM and spectro-ptychography is further illustrated in Figure

S14. Crystallinity was investigated using SAED (Selected Area Electron Diffraction). Figure 6b and 6c present high resolution TEM images of a typical cluster at the surface of the Ti-doped hematite nanoparticles. The diffractograms were obtained using FFT and used for structural indexation (Figure 6d). More details are given in Figure S13. We can identify diffraction spots arising from hematite and pseudo-brookite structures, with their typical d-spacing at ~0.275 nm and ~0.355 nm, respectively.

We may attempt here an explanation for the mechanism of pseudo-brookite $Fe_2TiO_5$ formation, even though a proper structural *in situ* investigation was not conducted during this study. We start our explanation following the same arguments developed elsewhere,[15] hypothesizing the existence of two kinds of Ti bonds in the bulk of the akaganeite iron oxyhydroxide structure formed after the ACG process. One is substituting Fe in Fe-$O_6$ octahedra and the other is bonded with the Cl from the akaganeite tunnel structure. This assumption allows to explain the Ti segregation toward the surface leading to Ti-rich phase upon annealing. Indeed, when increasing the temperature, the Cl bonded with Ti will easily migrate toward the surface of the hematite resulting in Ti surface segregation. Such segregation was reported equally for Ti-doped hematite obtained by different synthesis methods,[23] but in the absence of adapted spectromicroscopy techniques, it was supposed that the Ti-rich phase is $TiO_2$. In our case the Ti-rich phase present Ti-$O_6$ coordination, but different in terms of interatomic distances and angles of the octahedra from those constituting the Ti-doped hematite structure (assimilated to an ilmenite structure). This idea is supported by the Ti $L_{2,3}$ absorption spectral features (*i.e.* shift toward higher energies of the Ti $L_3$ $t_{2g}$ and asymmetry of the Ti $L_3$ $e_g$ peak). Indeed, the $Fe_2TiO_5$ can be expressed as a phase aggregation: $Fe_2TiO_5 = Fe_2O_3 + TiO_2$. Interestingly, this aggregation occurs only when annealing under Nitrogen, the air annealing producing only a Ti substitution gradient without pseudo-brookite phase formation in that case.[15] Most probably the presence of Oxygen stabilizes a homogeneous phase at the surface of the Ti-doped hematite that can be assimilated to ilmenite, while Nitrogen promotes phase separation between Ti-doped hematite ($Fe_{2-x}Ti_xO_3, with\ x \leq 1$) and pseudo-brookite.

## 3. CONCLUSIONS

In summary, we demonstrate enhanced photoelectrochemical performances by simply annealing Ti-doped hematite photoanodes grown by ACG under Nitrogen compared to the commonly used air annealing. We present comparatively the results obtained from two samples: one annealed under Nitrogen and the other in air. Photocurrent measurements and impedance spectroscopy reveal that the main mechanism responsible for the enhanced activity is the increased charge transfer rate at interface with the electrolyte, promoted by higher oxidation energy surface states. Interestingly, oxygen vacancies play a secondary role in this case, as demonstrated by the comparison of S1 and S2 samples obtained with very low Ti doping. Combined S-XPEEM, STXM and ptychography spectromicroscopies, employing state of the art instruments dedicated to synchrotron radiation 26haracterization, unravelled the origin of these surface states. They are related to the formation of pseudo-brookite clusters at the surface of Ti-doped hematite nanorods. Using the particular XPEEM geometry we access both surface and bulk chemical and coordination properties at nanoscale. We could thus evidence modifications of O K-edge spectra directly linked to the formation of surface Ti-rich phase promoting surface states in the unoccupied DOS. We report here, for the first time , spectro-ptychography measurements at Ti $L_{2,3}$ absorption edges. We demonstrate how very strong chemical coordination sensitivity allows probing surface clusters as small as 10 nm. DFT calculations and complementary STEM measurements consolidated our findings, confirming the formation of pseudo-brookite clusters at the surface of Ti-doped hematite nanoparticles. These clusters are the origin of the enhanced PEC activity. Finally, we show that low-tech approach allows escalating solar water splitting efficiency of photoanodes obtained from earth abundant material.

## 4. EXPERIMENTAL SECTION

4.1 SAMPLES PREPARATION

Ti-substituted $Fe_2O_3$ nanorod photoanodes were grown on $F:SnO_2$ (FTO) substrates following a process in two steps. First, Ti-doped akaganeite (Ti:β-FeOOH) films were elaborated by aqueous chemical growth (ACG) as detailed elsewhere.[15] Second, the akaganeite phase was converted in hematite by 600°C annealing in air or in a furnace quartz tube in presence of $N_2$ flow (3 l/h). The annealing ramp is presented in Figure S15. We used an aqueous solution with pH value of 1.4. The hydrothermal activation temperature during the growth was 95 °C and the deposition time of 24 h. These growth parameters allowed us to obtain brush-like nanorods, with nanorods perpendicularly oriented on the FTO substrate. Nanorods have typical 50 nm width and 200 nm length. Following this procedure, we obtained 9% average Ti substitution level, with increasing gradient from the FTO substrate toward the tip of the nanorods .[15] For XPEEM and STXM measurements the samples were prepared collecting Ti-doped iron oxyhydroxide (FeOOH) powder from the flasks used for the ACG process and left several days at room temperature for dehydration. The FeOOH powder was then annealed at 600°C either in air or under $N_2$ controlled flow conditions to obtain the hematite phase. Hematite powder was drop-casted using a solvent carrier of 1:1 isopropanol in ultra-pure water, either on gold-plated Si wafers for S-XPEEM or on location-tagged holey TEM SiN membranes, for STXM, spectro-ptychography, and STEM measurements. No signs of surface sample alteration were observed due to the use of isopropanol – water mixture carrier. Such alterations would have hindered especially the data recorded in XPEEM, strongly sensitive to surface topmost atomic layers. In addition, the PEC results are fully coherent with the XPEEM ones.

4.2. PHOTOCURRENT MEASUREMENT

Stabilized values of the photocurrent were measured using a dedicated setup equipped with an UV-vis light source (Newport 1000 W Xe arc lamp), a three-electrode cell allowing front illumination of the photoanode and a Princeton Applied Research 263A potentiostat. All measurements were performed at room temperature using 0.1 M NaOH as electrolyte. The samples are mounted as working electrode, a platinum wire is used as counter electrode, and Ag/AgCl electrode as reference.

4.3. ELECTROCHEMICAL IMPEDANCE SPECTROSCOPY (EIS) MEASUREMENTS

A CompactStat IVIUM potentiostat connected to a three-electrode photoelectrochemical H-cell (Redox.me), where hematite samples are mounted as working electrode, allowed to measure the impedance, $Z$, as a function of the frequency in the 10 Hz – 900 kHz range and of the applied potential varying between 0.7 – 1.6 V *vs.* RHE. The complex impedance:

$$Z = \frac{V(\omega)}{I(\omega)} = Z_{re} + jZ_{img}$$

contains information on the RC-like equivalent circuit corresponding to the photoanode-electrolyte interface. Nyquist plots expressing $|Z_{img}|$ *vs.* $(Z_{re}, f)$ and Mott-Schottky analysis were obtained by modeling the data using the IVIUM software applied to the equivalent RC circuit presented in Figure S3. In this manner we estimated flat band potential and carrier concentration (see Supporting Information for more details).

4.4. X-RAYS SPECTROMICROSCOPY

Spectromicroscopy combines milli-eV spectral and nanometer spatial resolutions allowing to address nanoscale physical and chemical properties. All data presented hereafter were obtained using two microscopes (XPEEM and STXM) available at the HERMES beamline[61] from the French SOLEIL synchrotron facility. Combining these techniques is of a particular interest when seeking surface/bulk information with very good spectral and spatial resolutions. Both microscopes use NEXAFS (Near Edge X-ray Absorption Fine Structure) as main contrast mechanism.

4.4.1. SHADOW X-RAY PHOTO-EMISSION ELECTRON MICROSCOPY (S-XPEEM)

We employed XPEEM[62] to probe surface chemistry and electronic structure of the samples at nanoscale. XPEEM is strongly surface selective. Indeed, due to the very short electron mean free path, $\lambda_e$, XPEEM, likewise XPS, is "blind" for photoelectrons generated deeper than ~3 nm ($3 \cdot \lambda_e$) from the surface.[63] Nevertheless, the specific XPEEM apparatus geometry, *i.e.* 16° incidence angle of the X-rays with respect to the sample surface plane, allow accessing equally bulk information, gathered from the sample shadow projected by the X-rays on the substrate. A detailed schematic is given in Figure S7. Pioneered by Kimling et al.,[64] the shadow X-rays photo-emission electron microscopy (S-XPEEM) was mainly employed by other groups to unravel complex buried magnetic structures.

4.4.2. SCANNING TRANSMISSION X-RAYS MICROSCOPY (STXM)

STXM is well established technique[65] in a raster scheme where each pixel value obeys the Beer-Lambert law, $I = I_0 \cdot e^{-\mu(E) \cdot \rho \cdot d}$, with $I$ the transmitted intensity, $I_0$ the direct X-ray beam intensity, $\mu(E)$ the energy dependent mass attenuation coefficient, $\rho$ the mass density, and $d$ the thickness of the sample. It is thus convenient to quantify material parameters in optical density (OD) units, measured as $\log(I_0/I)$. In addition, $\mu(E)$ is accessed recording hyperspectral data (energy stacks), where the energy of the impinging X-rays is tuned, and an image is recorded at each energy point. One can thus define speciation maps where each spatially resolved pixel carry well defined spectral information.

4.4.3. SPECTRO-PTYCHOGRAPHY

Spectro-ptychography is performed at HERMES beamline at SOLEIL synchrotron, replacing the standardly used photomultiplier tube with a sCMOS camera.[57] Scattering images are recorded at each (X,Y) sample coordinate and the final real space image is obtained using the ptychographic reconstruction algorithm implemented in the PyNX suite.[66] In addition, energy stacks are obtained in spectro-ptychography mode, recording data and then reconstructing ptychography images at various energy points. Amplitude and phase reconstructed images are reassembled in stacks and treated as STXM energy stacks, *i.e.* drift correction and SVD. The acquisition control is ensured using Python scripts through the PyTango binding allowing to interact with the sCMOS camera Tango device. An external clock gating scheme is used to trigger the camera, such as to obtain the synchronization between the sample stage positioning and the camera exposure. Two timing schemes are employed: fast one, typically at 20 fps, and a slow one, typically at 10 or 5 fps. The fast scheme is used simply to replace the normal PMT, STXM images being reconstructed in real time integrating the bright field of the images. This is used for alignment, sample navigation and focusing, likewise in standard STXM experiments. The slow scheme is used when recording data for PyNX ptychographic reconstructions. Here we used typically $2.5 \times 2.5$ µm² scanning sample areas and a defocused 1 µm X-ray beam spot size, obtained displacing the Fresnel zone plate upstream.

4.5. TRANSMISSION ELECTRON MICROSCOPY

The electron microscopy analyses were performed using a JEOL 2100 FEG S/TEM microscope operating at 200 kV and equipped with a probe spherical aberration corrector. For the acquisition of HAADF images in the STEM mode, a spot size of 0.13 nm, a current density of 140 pA, a camera focal length of 8 cm, which corresponds to inner and outer diameters of the annular detector of 73 and 194 mrad, have been used. The elemental maps were also recorded in the STEM mode by EDXS using a silicon drift detector (SDD) with a sensor size of 60 mm².

4.6. DENSITY FUNCTIONAL THEORY CALCULATIONS

Electronic structure calculations of ilmenite and pseudo-brookite phases have been performed using the very efficient localized-orbital basis set DFT code Fireball.[67,68] In this method, the self-consistency is achieved over the occupation numbers through the Harris functional[69] and LDA exchange-correlation energy is calculated using the efficient multi-center weighted exchange correlation density approximation (McWEDA).[70] This method has already been used for the study of molecular adsorption on $Fe_3O_4$ surfaces in good agreement with experimental determination.[71]

SUPPORTING INFORMATION

1 – Fast sweep and stabilized photocurrent measurements; analysis of electrochemical impedance spectra: Nyquist plots, equivalent circuit, Mott-Schottky plots, flat band potential, band bending; shadow-XPEEM: surface *vs*. bulk contributions, specific geometry, normalization of spectral contributions, model structures for ilmenite and pseudo-brookite; spectro-ptychography: experimental conditions, amplitude and phase signals; electron microscopy: STEM and SAED indexation, crystallographic structure; correlation STXM – STEM; annealing ramp (PDF)

2 – S-XPEEM hyperspectral data (movie) of the sample annealed under Nitrogen (S1 sample) (GIF)

3 – Spectro-ptychography hyperspectral data (movie) of the sample annealed under Nitrogen (S1 sample) (GIF)

AUTHOR CONTRIBUTIONS

Conceptualization, S. Stanescu and D. Stanescu; data curation, S. Stanescu, D. Stanescu, T. Alun, D. Ihiawakrim and Y. J. Dappe; formal analysis, S. Stanescu, D. Stanescu, D. Ihiawakrim, O. Ersen and Y. J. Dappe; funding acquisition, S. Stanescu and D. Stanescu; investigation, S. Stanescu, T. Alun, D. Ihiawakrim and D. Stanescu; methodology, S. Stanescu and D. Stanescu; project administration, S. Stanescu and D. Stanescu; spectro-ptychography acquisition software, S. Stanescu; supervision, S. Stanescu; validation, S. Stanescu, D. Stanescu, Y. J . Dappe and O. Ersen; writing – original draft, S. Stanescu and D. Stanescu; and writing – review & editing, all.

ACKNOWLEDGEMENTS

This research was funded, in whole or in part, by a public grant overseen by the French National Research Agency as part of the "Investissements d'Avenir" program (Labex NanoSaclay, reference ANR-10-LABX-0035). A CC-BY public copyright license has been applied by the authors to the present document and will be applied to all subsequent versions up to the Author Accepted Manuscript arising from this submission, in accordance with the grant's open access conditions. D.S. thanks to Anne Forget for support using the chemistry laboratory facility at CEA-Saclay/IRAMIS/SPEC. S.S. acknowledge help from Nicolas Mille for implementation of the spectro-ptychography at the HERMES beamline. Synchrotron SOLEIL facility is acknowledged for providing beamtime under project numbers 99190036 and 20210794.

# Enhancement of the Solar Water Splitting Efficiency Mediated by Surface Segregation in Ti-doped Hematite Nanorods

*Stefan Stanescu[a]\*, Théo Alun[b], Yannick J. Dappe[b], Dris Ihiawakrim[c], Ovidiu Ersen[c], and Dana Stanescu[b]\**

[a] Synchrotron SOLEIL, L'Orme des Merisiers, Départementale 128, 91190 Saint-Aubin, France
E-mail: stefan.stanescu@synchrotron-soleil.fr

[b] SPEC, CEA, CNRS, Université Paris-Saclay, CEA Saclay 91191 Gif-sur-Yvette Cedex, France
E-mail: dana.stanescu@cea.fr

[c] Institut de Physique et Chimie des Matériaux de Strasbourg (IPCMS), CNRS UMR 7504, 23 rue du Loess, BP43, 67034 Strasbourg, France

I. Fast sweep and stabilized photocurrent measurements

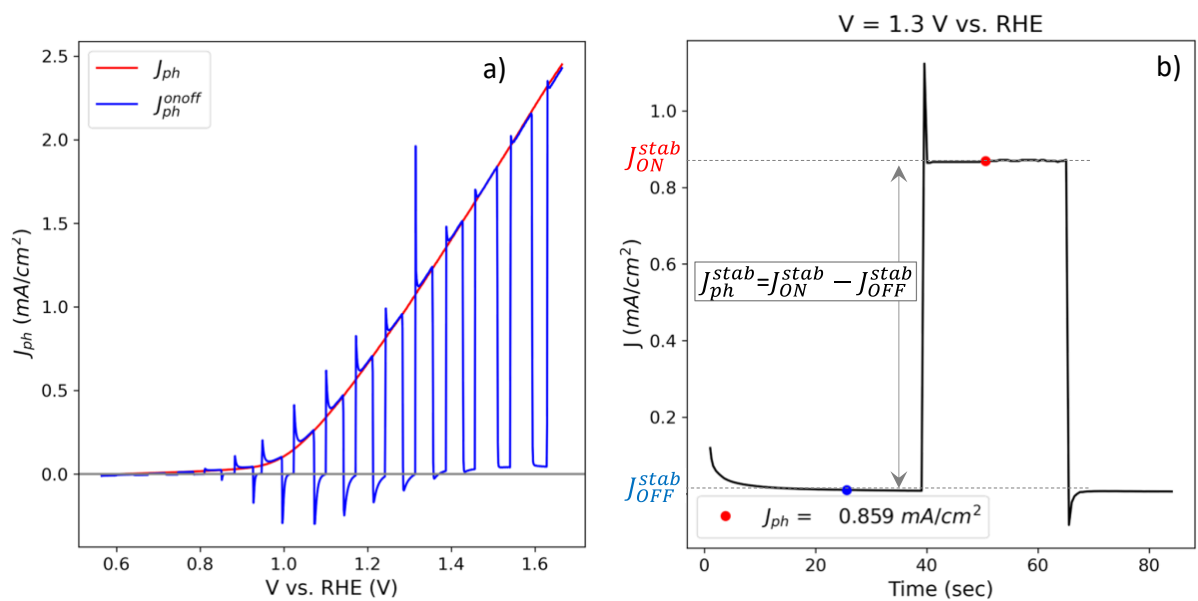

**Figure S1:** Photocurrent density measurement. a) fast sweep photo-voltammetry where the applied voltage is swept at a rate of 50 mV/s; (b) stabilized photocurrent measurement where the current is measured as a function of time at constant voltage. In these figures one presents the photocurrent measured for S1 sample.

II. Analysis of electrochemical impedance spectra : Nyquist plots, equivalent circuit, Mott-Schottky plots, flat band potential and band bending $R_1$ ($R_2$) and $Q_1$ ($Q_2$) are the resistance respective the constant-phase-element (CPE) associated to the semiconductor photoanode (Helmholtz layer). $R_3$ represents a serial resistance due to contact on the sample in general. CPEs are introduced to model imperfect dielectrics, both for semiconductor (Q1) and Helmholtz layer (Q2). These replace ideal capacitors of capacitance C1 and C2 respectively, the corresponding impedances, $Z_1^{CPE}$ and $Z_2^{CPE}$ can be calculated using the relation:

$$Z_i^{CPE} = \frac{1}{(j\omega)^{N_i} \cdot Q_i}; \quad i = 1,2; \quad \omega = 2\pi f \tag{S1}$$

where $f$ is the frequency, $Q_i$ and $N_i$ are frequency independent quantities at a given temperature, and $N_i$ can take values between 0 and 1, $N_i = 1$ for an ideal capacitor.

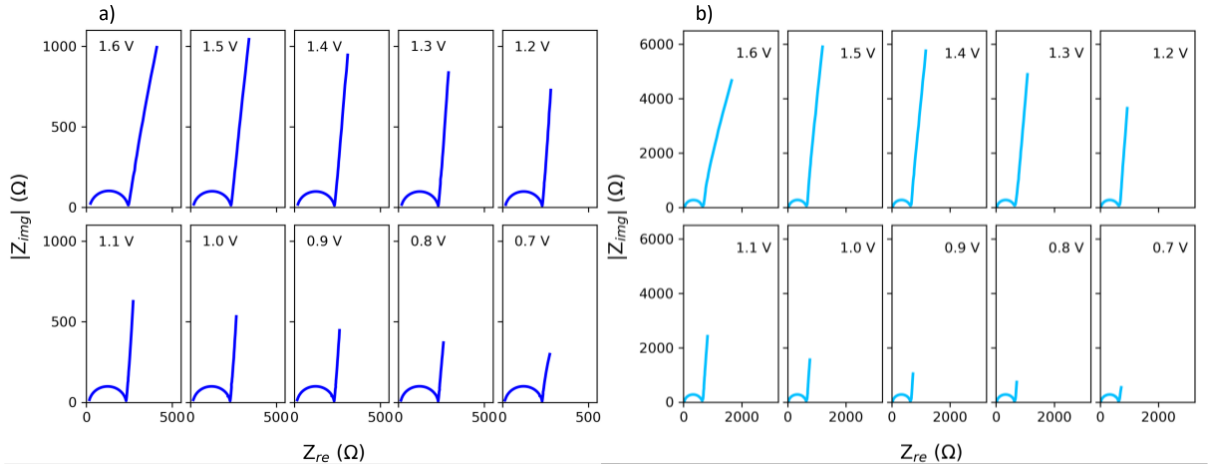

FIGURE S2: NYQUIST PLOTS REPRESENTING IMAGINARY PART – $|Z_{IMG}|$ – AS FUNCTION OF THE REAL PART – $Z_{RE}$ FOR S1 (A) AND S2 (B) SAMPLES.

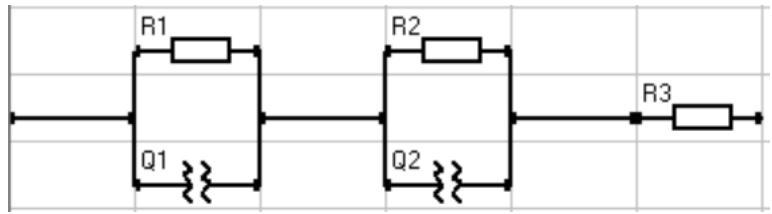

FIGURE S3: EQUIVALENT CIRCUIT USED TO FIT EIS SPECTRA

Frequency dependent complex capacitances, $C_i$, can be calculated from:

$$Z_i^{CPE} = \frac{1}{(j\omega)^{N_i} \cdot Q_i} = \frac{1}{j\omega C_i} \tag{S2}$$

using the formula:

$$\tilde{C}_i(\omega) = Q_i(j\omega)^{N_i - 1} \tag{S3}$$

Capacitance values reported in this paper, C, (Figure 2 and Figure S5) correspond to the magnitude of the complex capacitance of the semiconductor, $\tilde{C}_1(\omega)$ :

$$C = \sqrt{Re\left(\tilde{C}_1(\omega)\right)^2 + Im\left(\tilde{C}_1(\omega)\right)^2} \tag{S4}$$

The variation with the applied voltage of parameters (Ri, Qi and Ni ) for both S1 and S2 samples, as resulted from the fit procedure using IVIUM software, are presented in Figure S4.

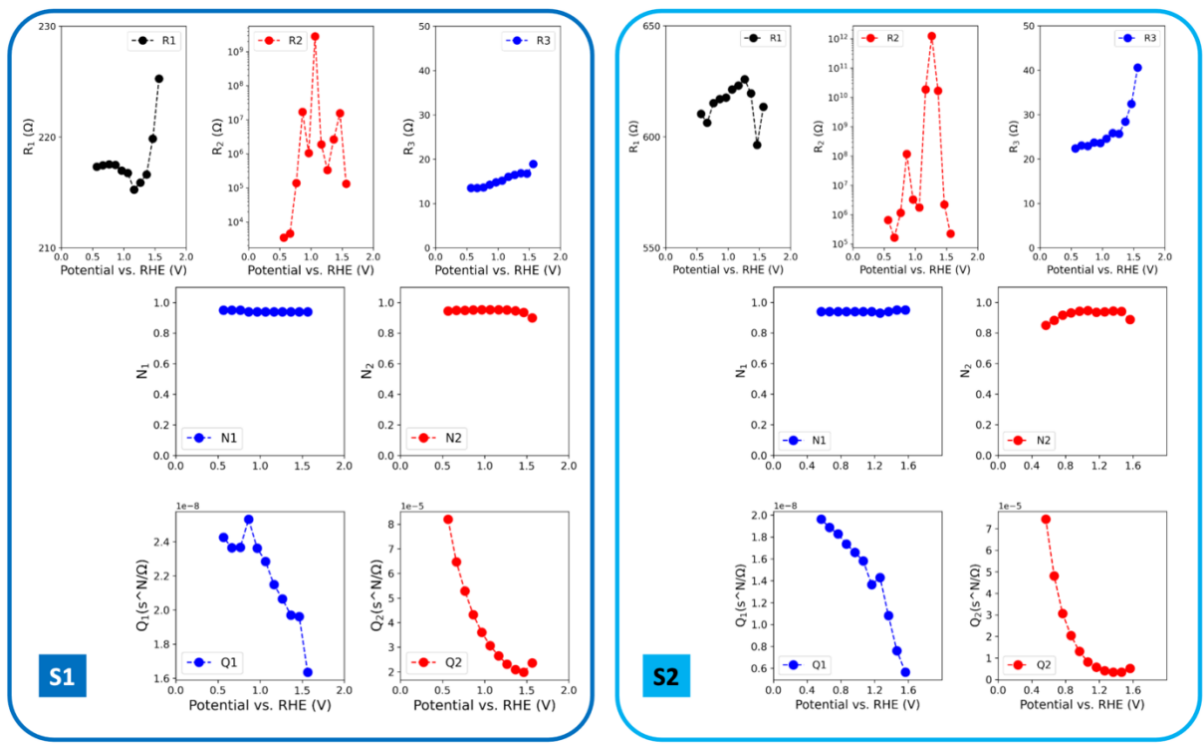

Figure S4: Fit parameters ($R_i$, $Q_i$ and $N_i$) for both S1 and S2 samples.

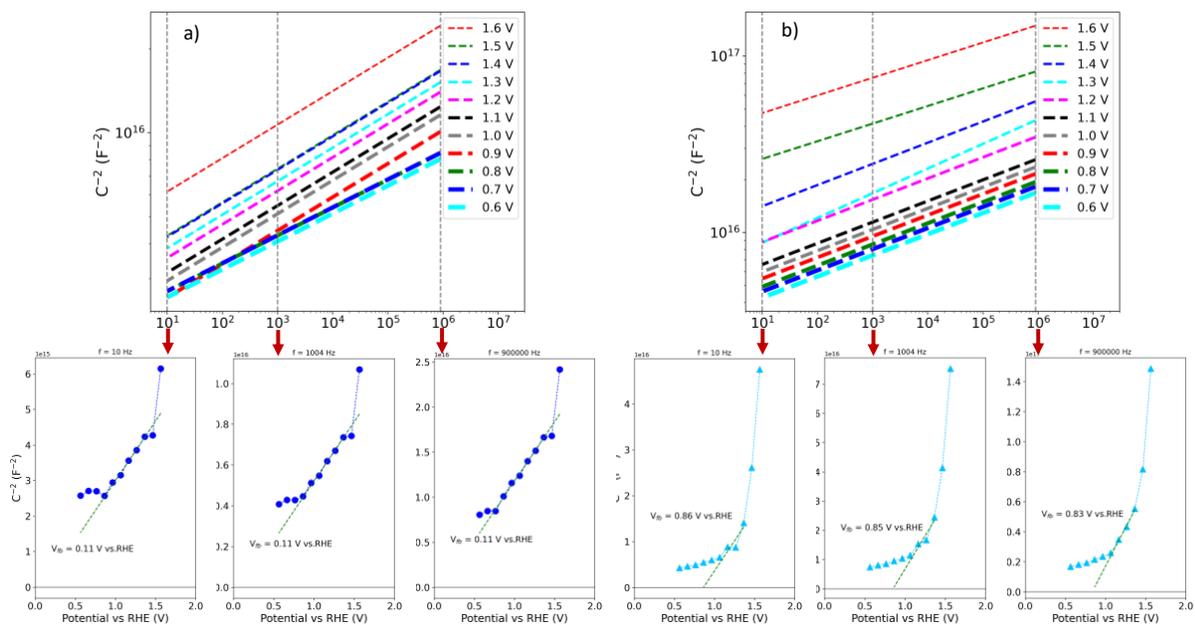

**FIGURE S5**: FREQUENCY DISPERSION OF $C^{-2}$ VALUES FOR POTENTIALS BETWEEN (0.6 – 1.6) V FOR BOTH S1 (A) AND S2 (B). FROM MOTT-SCHOTTKY CURVES AND LINEAR FITS WE DEDUCED FREQUENCY INDEPENDENT FLAT-BAND VALUES, ~ 0.1 V FOR S1 AND ~ 0.8 V FOR S2.

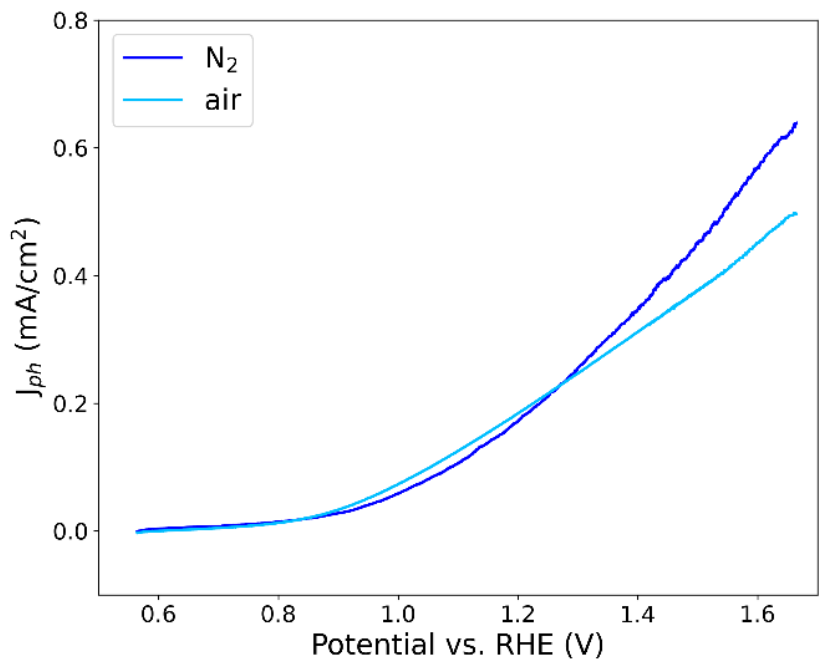

**Figure S6**: Comparative photocurrent measurements realized on very low doped Ti samples (0.01 mass % TiCl$_3$), that can be assimilated to pristine hematite, annealed under Nitrogen and air. There is no visible difference between the two situations at 1.23 V vs. RHE, suggesting a low contribution related to the presence of oxygen vacancies.

## III. Shadow XPEEM – surface/bulk contributions

The particular XPEEM geometry, *i.e.* X-rays describe 16° incidence angle with the sample surface plane, allows useful surface/bulk discrimination. Especially when using strongly 3D-shaped materials, this leads to a large shadow projection carrying bulk information. Figure S7 shows the geometry of the S-XPEEM.

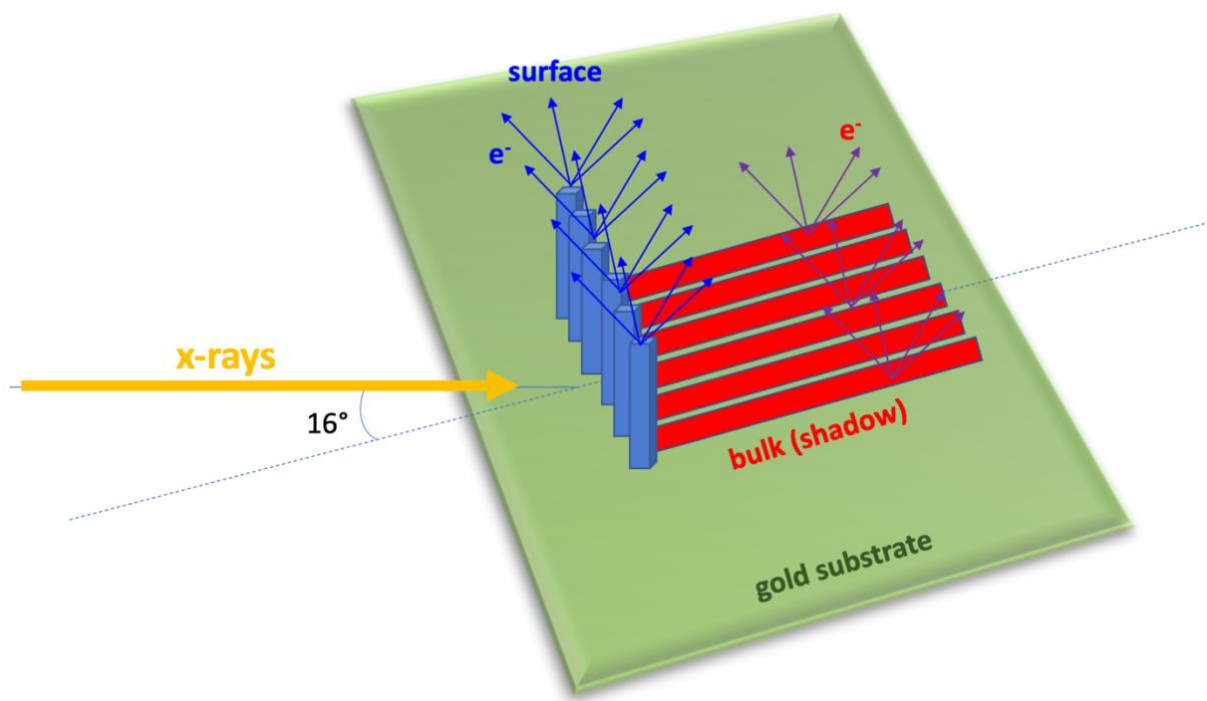

**Figure S7.** Shadow XPEEM geometry: X-rays (orange) describe 16° incidence with the sample plane. Two regions are distinguishable: the surface (Blue) and the bulk (Red).

The incident X-rays will generate photoelectrons through the photoelectric effect. When tuning the X-rays energy across an absorption edge, the intensity of the photoemission signal will strongly increase, being proportional to the absorption coefficient, µ. A strong contrast will appear thus at the absorption resonance between different regions in the S-XPEEM images. At the end, in our case, the spectral sensitivity driving the contrast mechanism is strictly related to the NEXAFS (*Near Edge X-ray Absorption Fine Structure*) sensitivity. Surficial sample regions, directly interacting with the incoming X-rays, will lead to the blue signal in **Figure S8a**. In the same manner, the green signal from the gold-plated substrate is obtained, exempted of any particular spectral features at the targeted absorption edges. We remark several "kinks" appearing both in the blue and green regions, related thus clearly with direct beam features.

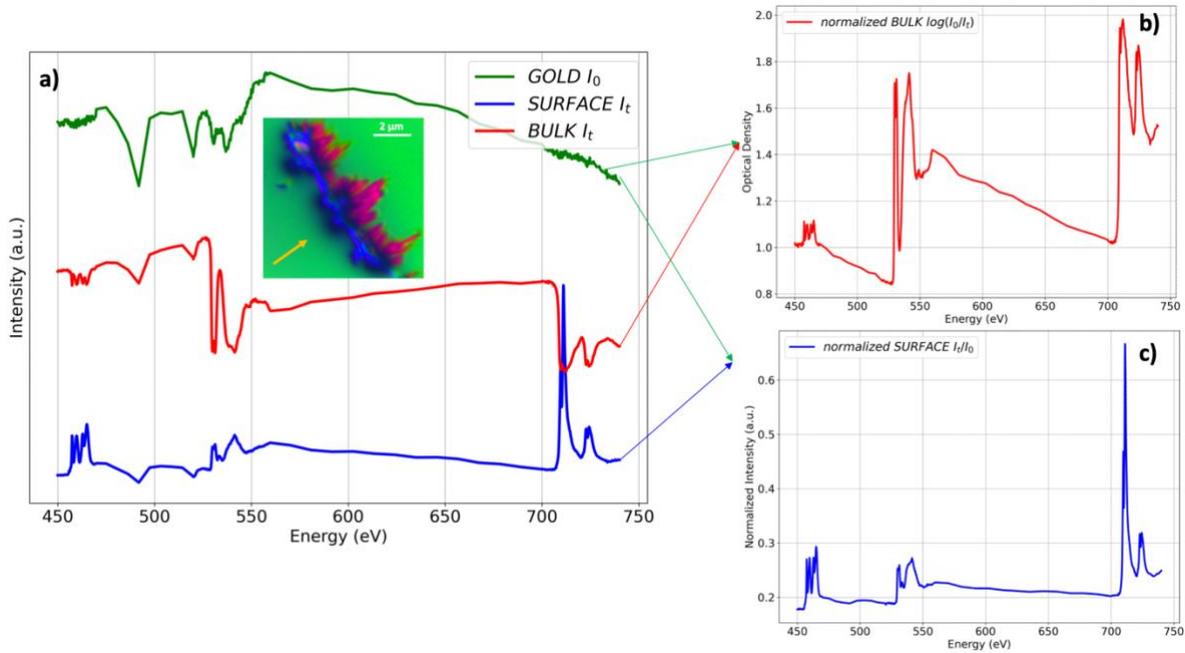

**Figure S8.** S-XPEEM signals normalization. Different normalisation from shadow (Red) and surface (Blue) contributions from the S1 sample.

The signal recorded from the substrate (green) is used as $I_0$ to normalize the surface and bulk spectra, as detailed in the following. First, to understand the red (shadow) signal in Figure S8a, the gold-plated substrate has to be seen as being part of the detection scheme. Indeed, we perform in this case an X-ray transmission experiment, the X-rays being transmitted through the sample and then detected using the gold substrate as photoelectron emitter, likewise scintillators are used in many standard detection schemes to transform X-rays in visible photons. Thus, the red signal collected from this region is directly proportional to the transmitted X-rays intensity. Finally, the surface signal can be treated like TEY (Total Electron Yield) signal, and the normalized spectra (Figure S8b) is obtained as the ratio with the $I_0$ signal. The intensity of the resulting signal cannot be quantified directly. To the contrary, the bulk signal is treated as in transmission experiment and the optical density can be derived using the Beer-Lambert law:

$$OD = \log\left(\frac{I_0}{I}\right) = \mu \cdot \rho \cdot d \tag{S5}$$

with $\mu$, the mass attenuation/absorption coefficient, $\rho$, the density of the sample, and $d$, the thickness of the sample. At the Fe $L_{2,3}$ edges (700 – 750 eV), due to the strong X-ray absorption, the red signal is saturated for all samples we measured. In this spectral region the X-ray attenuation length is about 80-120 nm, the sample thickness is larger than 200 nm.

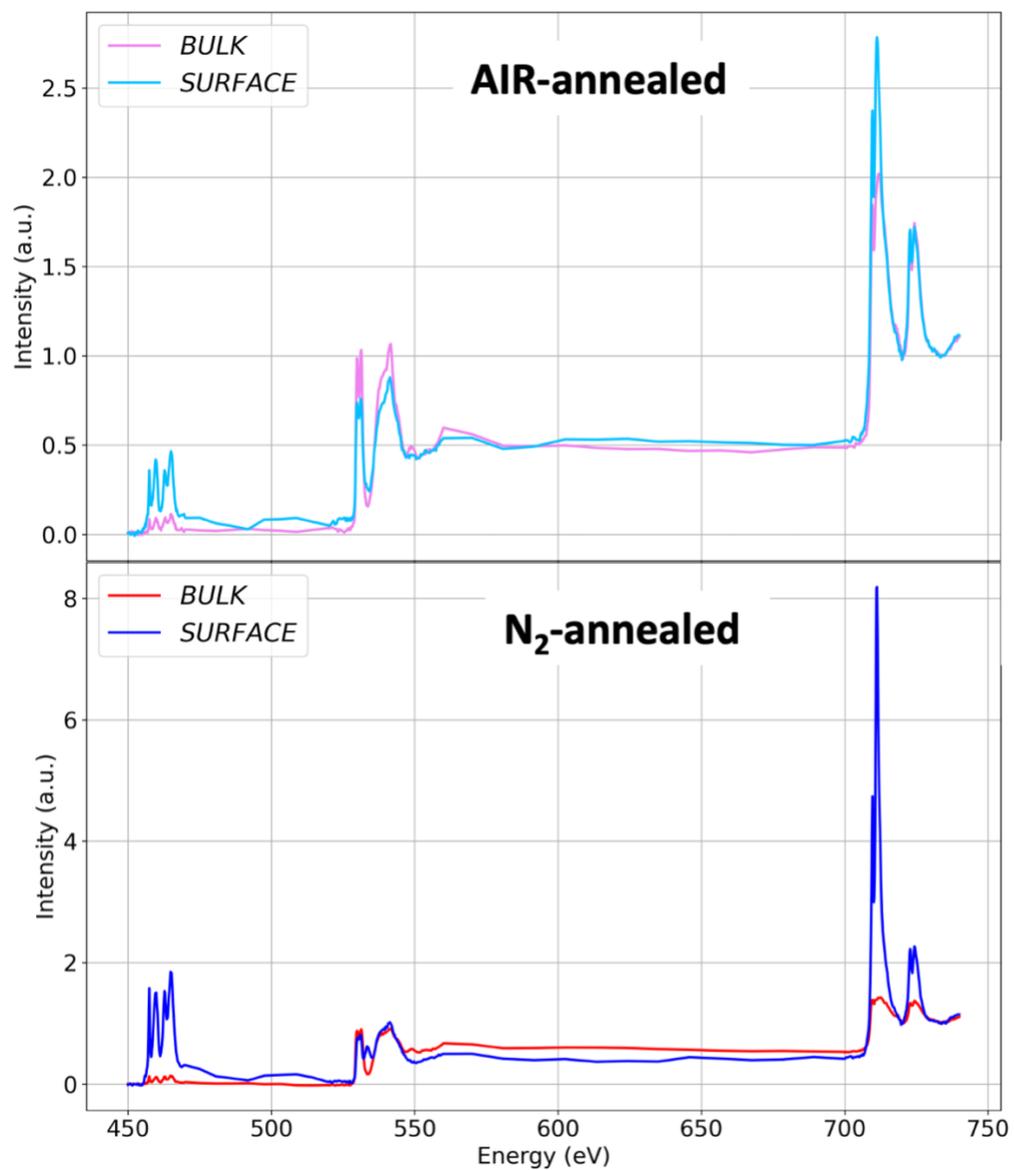

**Figure S9.** Bulk *vs*. surface NEXAFS spectra for S1 (bottom) and S2 (top) samples.

Hyperspectral S-XPEEM data for S1 sample, converted into animated GIF file, is shown as a separate file <S-XPEEM_S1.gif>.

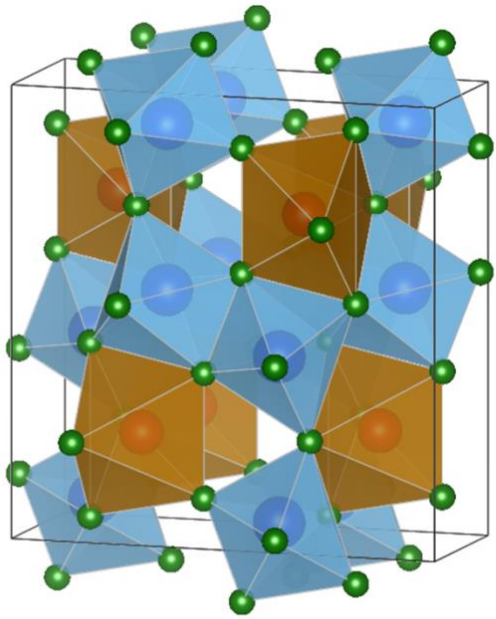
FeTi$_2$O$_5$

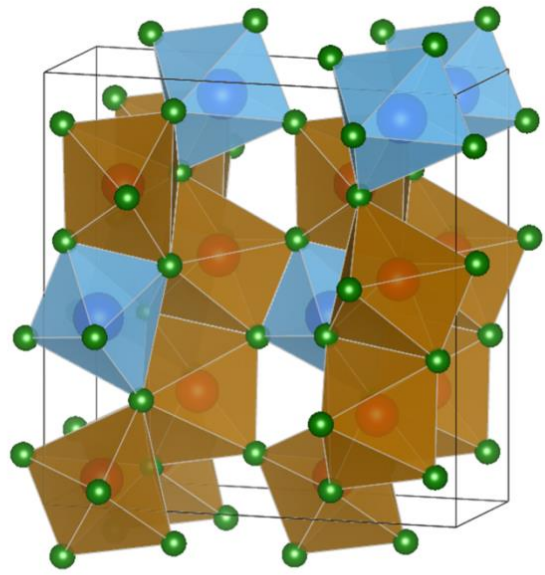
Fe$_2$TiO$_5$

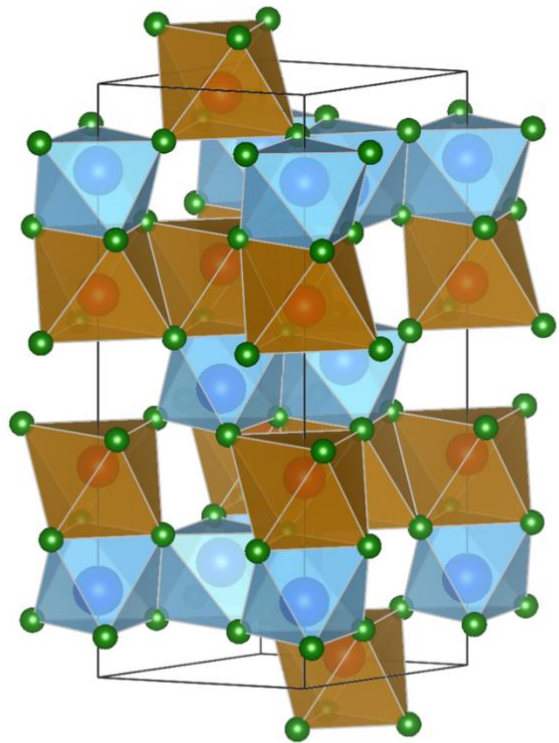
FeTiO$_3$

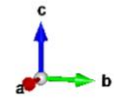

Figure S10. Iron titanate structures: pseudo-brookite ($Fe_{1+x}Ti_{2-x}O_5$ with x=0 and x=1) (top) and ilmenite (bottom). Iron ions are marked as red, titanium blue and oxygen green. Structures modeled using the Vesta software [1] and input CIF datafiles available from the Materials Project database [2].

## IV. Spectro-ptychography: some experimental conditions

As shown in the Figure S11, during the spectro-ptychography measurements we over-sampled the chosen sample area with at least 30 x 30 points, ensuring more than 92 % overlap between adjacent points, imperious criteria for the ptychographic reconstructions algorithms.

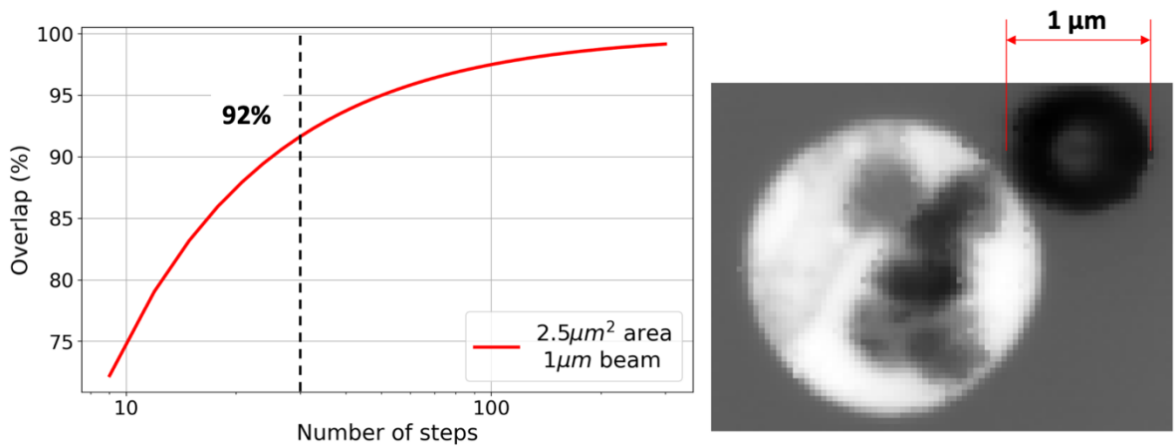

Figure S11. Left panel: overlap factor used for typical scanning area of 2.5 x 2.5 µm2 with defocused 1 µm beam spot size. Right panel: STXM image using the sCMOS camera detector showing the sample and the trace of the 1 µm X-ray beam.

In the same Figure S11, the right panel, we show an STXM image recorded using the sCMOS detector, in focus conditions, after having left the X-ray beam defocused at 1 µm for several hours such as to get visible carbon cracking on the SiN membrane. We can thus clearly see the X-rays beam footprint, *i.e.* 1 µm defocused X-ray beam spot, that can be directly compared with the location tagged holey TEM SiN membranes [3], presenting 2 µm holes, used for this study. Therefore, as it can be observed in Figure S11, right panel, it was handy to find self-supported sample regions used for all the STXM and spectro-ptychography measurements presented here. Using the strongly defocused beam, > 30 times compared to the in-focus beam size, allow scanning larger sample areas, with a good spatial resolution. Figure S12 presents an example of a larger scanned sample area, of 5 x 5 µm$^2$. For now, the only limitation for the scanned area is related with data size that can be processed by PyNX. For instance, the data processed in Figure S12 is composed of 7 files (one for each X-rays energy from 457.3 eV to 459.3 eV) of 13.4 Go each, corresponding to 50 x 50 diffraction images of 1700 x 1700 pixels and a pixel depth of 16 bits. Increasing the area imposes increasing the number of points per image to keep the same overlapping factor. It is interesting to remark the sensitivity of the approach, since it is possible to spatially separate regions exhibiting only 0.1 eV spectral shift.

Hyperspectral spectro-ptychography data across Ti and O absorption edges for S1 sample, converted into animated GIF file, is shown as separate file: < spectro-PTYCHO_S1.gif>.

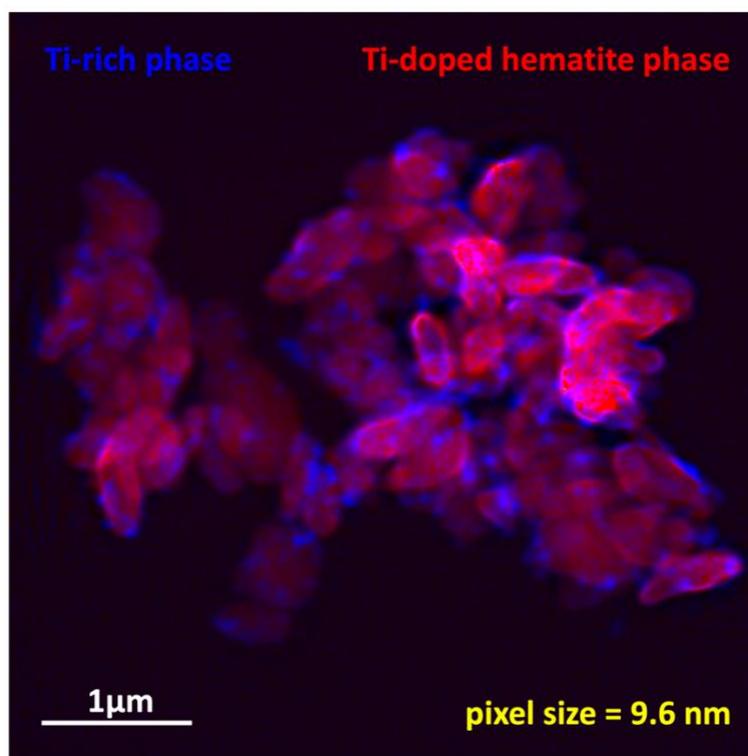

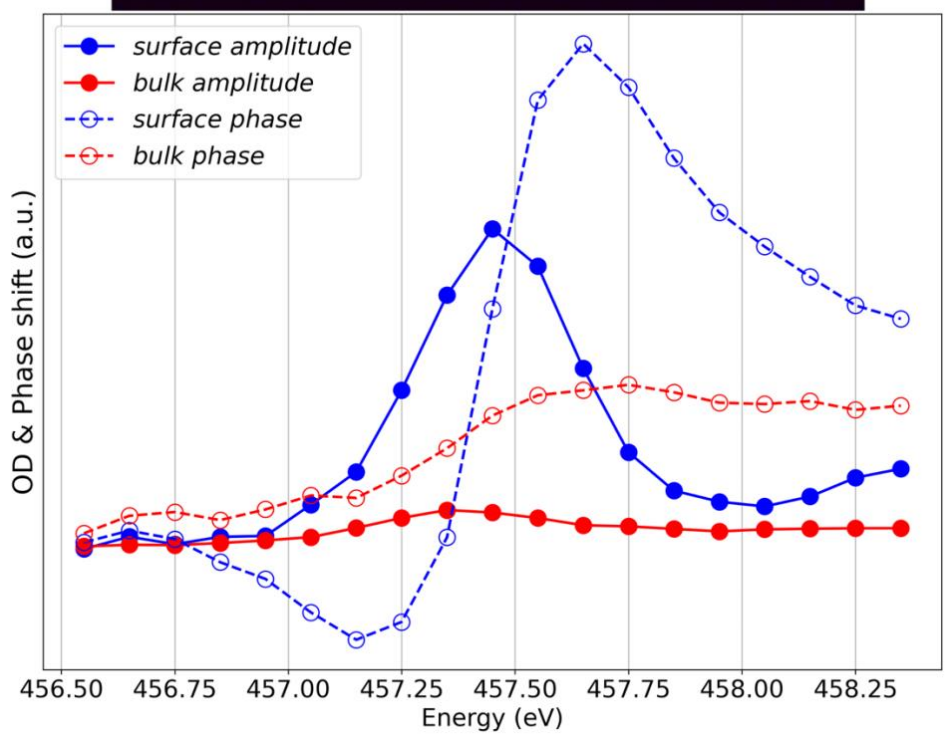

**Figure S12.** Ptychography data obtained on S1 sample. Top panel: ptychographic reconstructed image of 5 x 5 µm² sample area, obtained with 9.6 nm pixel size. Bottom panel: amplitude and phase signals extracted

from the ptychographic reconstructed hyperspectral data over 20 energies across the Ti $L_3$ $t_{2g}$. The phase shift signal from the surface Ti-rich phase dominates.

## V. STEM - SAED

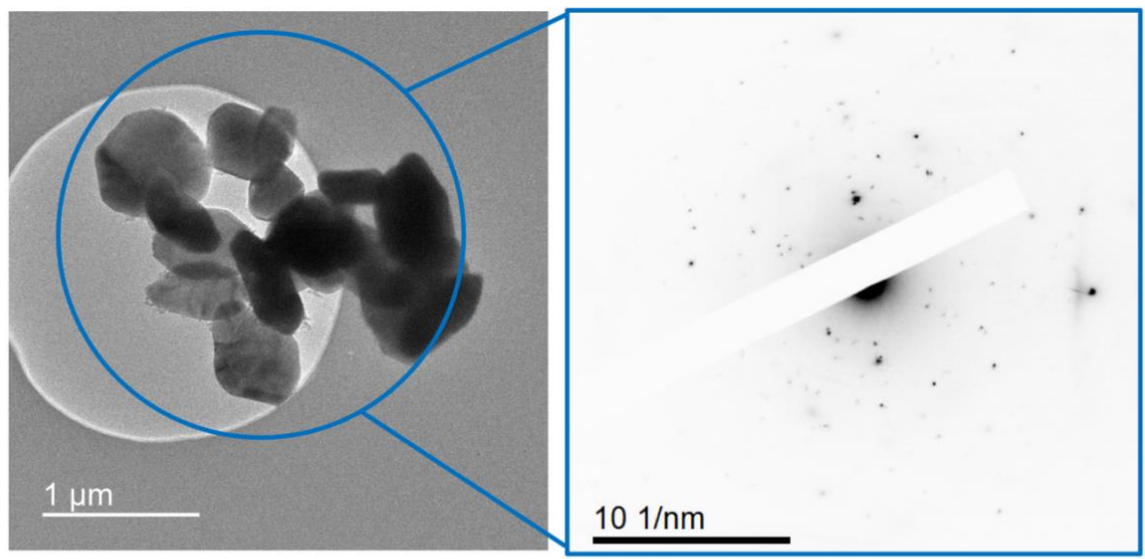

**Figure S13.** Selected area diffraction (SAED) structure determination from STEM. Top left image shows the Ti-doped hematite particles presenting small clusters at their surface. Selected region for the SAED is marked in blue and the associated diffractograms is shown in the top right panel. The identification is given in the bottom table where hematite, pseudo-brookite and ilmenite diffraction lines are indicated by colors.

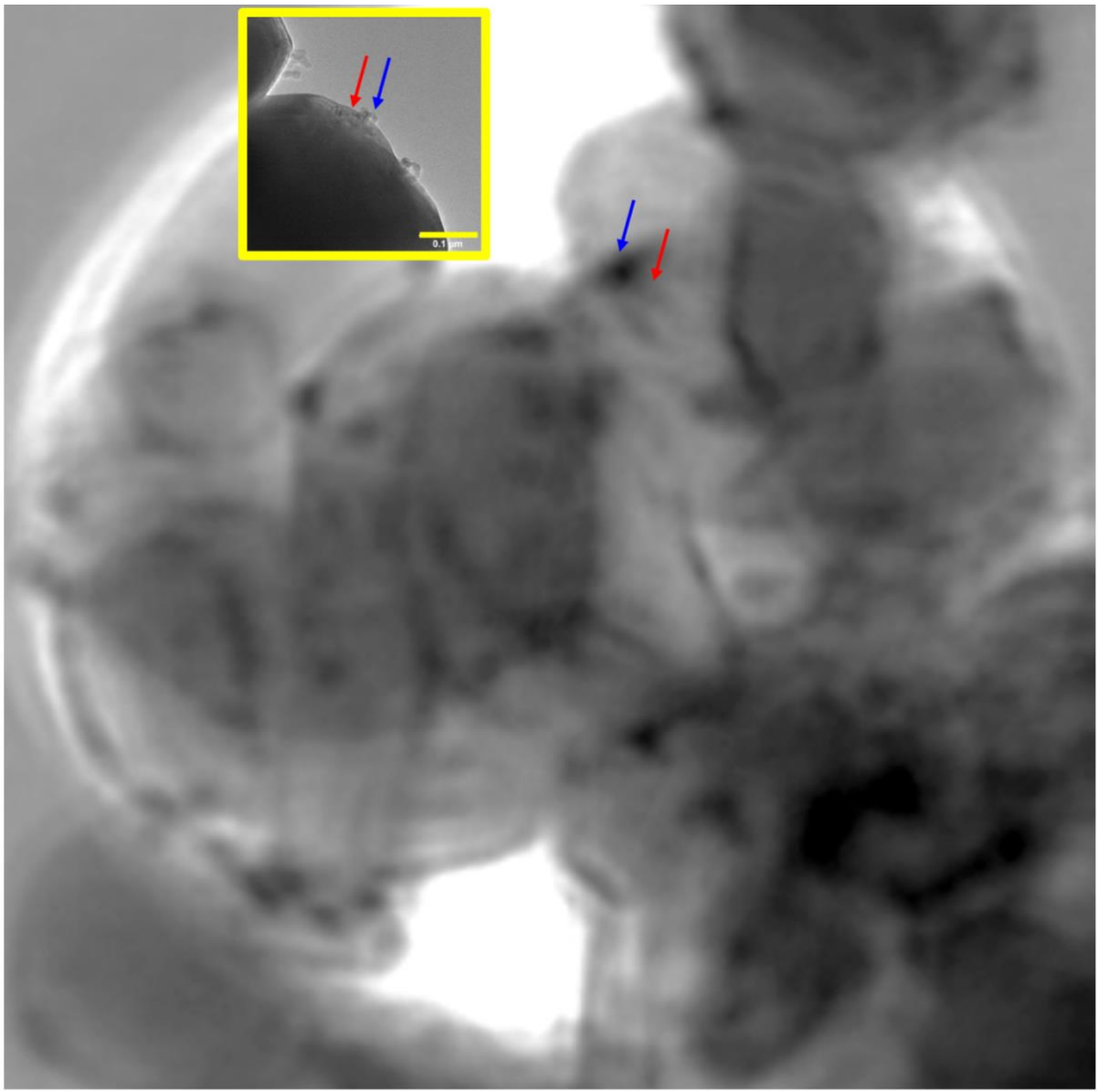

**Figure S14.** We compare here the probed length scales from spectro-ptychography and STEM (inserted yellow border figure). Amplitude image at 457.4 eV X-rays energy extracted from the spectro-ptychography hyperspectral data. Darker regions correspond to strongly absorbing pseudo-brookite clusters. The STEM (inserted yellow bordered) image scale was adapted to match the ptychography one. A common scale bar shown in STEM has 100 nm. Two regions can clearly be identified: first, marked with blue arrows, corresponds to columnar pseudo-brookite clusters. Second, marked with red arrows, presenting similar spectral character as the pseudo-brookite (absorbing more at the same resonant X-rays energy) and appearing clearly in the STEM image as a transition region between the pseudo-brookite and hematite. These are visible in the supplementary hyperspectral spectro-ptychography animated GIF file, <spectro-PTYCHO_S1.gif>.

## VI. Annealing ramp:

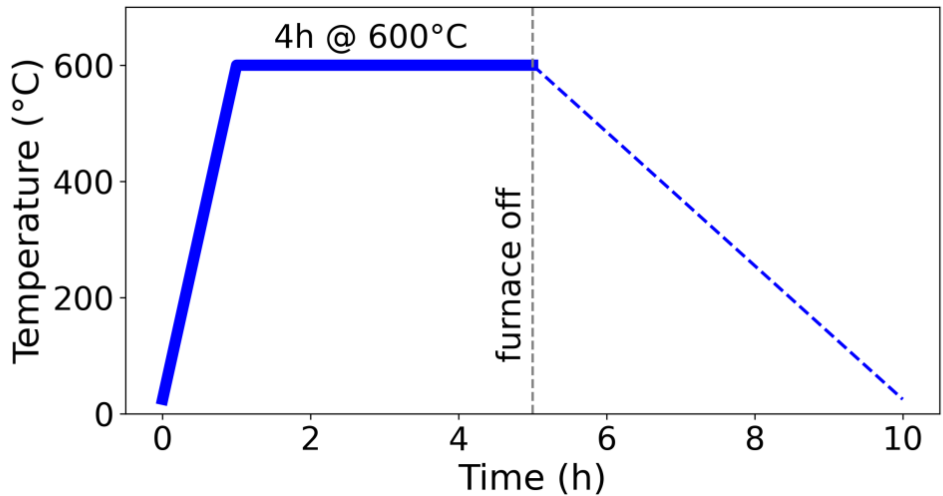

**FIGURE S15**: ANNEALING RAMP USED TO ANNEAL THE PHOTOANODES BOTH IN AIR AND NITROGEN.

This research was funded, in whole or in part, by a public grant overseen by the French National Research Agency as part of the "Investissements d'Avenir" program (Labex NanoSaclay, reference ANR-10-LABX-0035). A CC-BY public copyright license has been applied by the authors to the present document and will be applied to all subsequent versions up to the Author Accepted Manuscript arising from this submission, in accordance with the grant's open access conditions.